\documentclass{iopart}
\usepackage{graphicx}
\usepackage{hyperref}

\begin{document}
\title[Permanent magnetic lattices  for ultracold
atoms and quantum degenerate gases]{Permanent magnetic lattices  for
ultracold atoms and quantum degenerate gases}
\author{Saeed Ghanbari,
 Tien D Kieu, Andrei Sidorov and Peter Hannaford}
\address{Centre for Atom Optics and Ultrafast Spectroscopy and \\ ARC Centre of
Excellence for Quantum Atom Optics \\
 Swinburne University of Technology, Melbourne, Australia 3122}
\begin{abstract}
We propose the use of periodic arrays of permanent magnetic films
for producing magnetic lattices of microtraps for confining,
manipulating and controlling small clouds of ultracold atoms and
quantum degenerate gases.  Using analytical expressions and
numerical calculations we show that periodic arrays of magnetic films
can  produce one-dimensional (1D) and two-dimensional (2D)
magnetic lattices with non-zero potential minima, allowing ultracold
atoms to be trapped without losses due to spin flips.  In
particular, we show that two crossed layers of periodic arrays of
parallel rectangular magnets plus bias fields, or a single layer of
periodic arrays of square-shaped magnets with three different
thicknesses plus bias fields, can produce 2D magnetic lattices of
microtraps having non-zero potential minima and controllable trap
depth.  For arrays with micron-scale periodicity, the magnetic
microtraps can have very large trap depths  ($\sim$0.5 mK for the
realistic parameters chosen for the 2D lattice) and very tight
confinement.

\end{abstract}

\section{Introduction }
Periodic optical lattices produced by the interference of
intersecting laser beams have been used extensively in recent years
to confine, manipulate and control small clouds of ultracold atoms
and Bose-Einstein condensates~\cite{Bloch}. Such lattices are ideal
tools for performing fundamental quantum physics experiments such as
studies of low-dimensional quantum gases~\cite{Laburthe} and quantum
tunnelling experiments including the BEC superfluid to Mott
insulator quantum phase transition~\cite{Greiner}. Optical lattices
also have potential application in quantum information processing
since they may provide storage registers for qubits based on neutral
atoms~\cite{Calarco,Monroe}.

An alternative approach for producing periodic lattices for
ultracold atoms is to use the magnetic potentials of periodic arrays
of magnetic microtraps.  Simple, one-dimensional (1D) magnetic
lattices consisting of arrays of 2D traps or waveguides have been proposed~\cite{Hinds} and constructed using  current-carrying
wires~\cite{Gunther} or permanent magnetic
structures~\cite{Barb,SinclairCurtis,Sinclair} on `atom chips', and two-dimensional (2D) lattices of magnetic microtraps produced by
crossed arrays of current-carrying wires have been
proposed~\cite{Yin,Grabowski}.

We have recently developed technology for producing high-quality
magnetic microstructures based on permanent, perpendicularly
magnetised magneto-optical $Tb_6Gd_{10}Fe_{80}Co_4$ films~\cite{Wang}.  These magnetic
microstructures have been used to construct periodic grooved
magnetic mirrors for ultracold atoms~\cite{Wang,SidorovMcSch}, which
in the presence of bias magnetic fields can be transformed into a 1D
magnetic lattice of 2D traps, and to produce atom chips for
ultracold atoms and Bose-Einstein condensates~\cite{Hall}.  The
$Tb_6Gd_{10}Fe_{80}Co_4$ magneto-optical films exhibit excellent magnetic properties
for atom optics applications: they can be deposited with high
perpendicular magnetic anisotropy; they have excellent magnetic
homogeneity; and they can have high permanent magnetisation ($4\pi M_z\hspace{-.1cm}\sim\hspace{-.1cm}3.8\hspace{.1cm}kG$),
large coercivity ($H_c\hspace{-.1cm}\sim\hspace{-.1cm}3\hspace{.1cm}kOe$) and relatively high Curie
temperature ($\sim\hspace{-.05cm}300^\circ C$).

In this paper we propose the use of periodic arrays of permanent
magnetic films for producing magnetic lattices of microtraps for
confining, manipulating and controlling small clouds of ultracold
atoms and quantum degenerate gases.  Using analytical expressions
and numerical calculations we show that it is possible to produce 1D
and 2D permanent magnetic lattices with non-zero potential minima in
which ultracold atoms prepared in low magnetic field-seeking states
can be trapped without losses due to Majorana spin flips.  In
particular, we show that  two crossed separated
layers of periodic arrays of parallel rectangular magnets plus bias
magnetic fields, or a single layer of periodic arrays of
square-shaped magnets having three different thicknesses plus bias
fields, can produce 2D magnetic lattices of microtraps with non-zero
potential minima and large and controllable trap depth and high trap
frequency.

Magnetic lattices based on permanent magnetic films have potential
advantages over optical lattices or magnetic lattices based on
current-carrying wires that make them attractive for atom optics
applications and compact integrated devices.  They do not involve (high intensity) laser beams or
any beam alignment, and there is no light scattering or decoherence
due to spontaneous emission.  They can produce highly stable reproducible
potential wells with low technical noise.  They can be produced with
large trap depth and large magnetic field curvature, leading to very
high trap frequencies, without heat dissipation, in contrast to
current-carrying microwire devices. Using modern
microtechnology, permanent magnetic lattices may be fabricated with a wide
range of periods from about $100 \hspace{.1cm}\mu m$ down to about $1 \hspace{.1cm}\mu m$ and
they can in principle involve variable spacing between the lattice
sites or complex potential shapes at each lattice site.  Finally, in magnetic lattices,
only atoms in low magnetic field-seeking states are trapped and it should be
possible to perform radiofrequency evaporative cooling {\it in situ} in the lattice,
thereby allowing the study of very low temperature phenomena in lattices.

\section{Analytical expressions for infinite magnetic lattices }\label{sec2}
\subsection{ Single infinite periodic array of magnets with bias fields}\label{sec2.1}
\begin{figure}[tbp]
\begin{center}
$\begin{array}{ccc}
\hspace{1cm}
\includegraphics[angle=90,width=3.2cm,height=3.03cm]{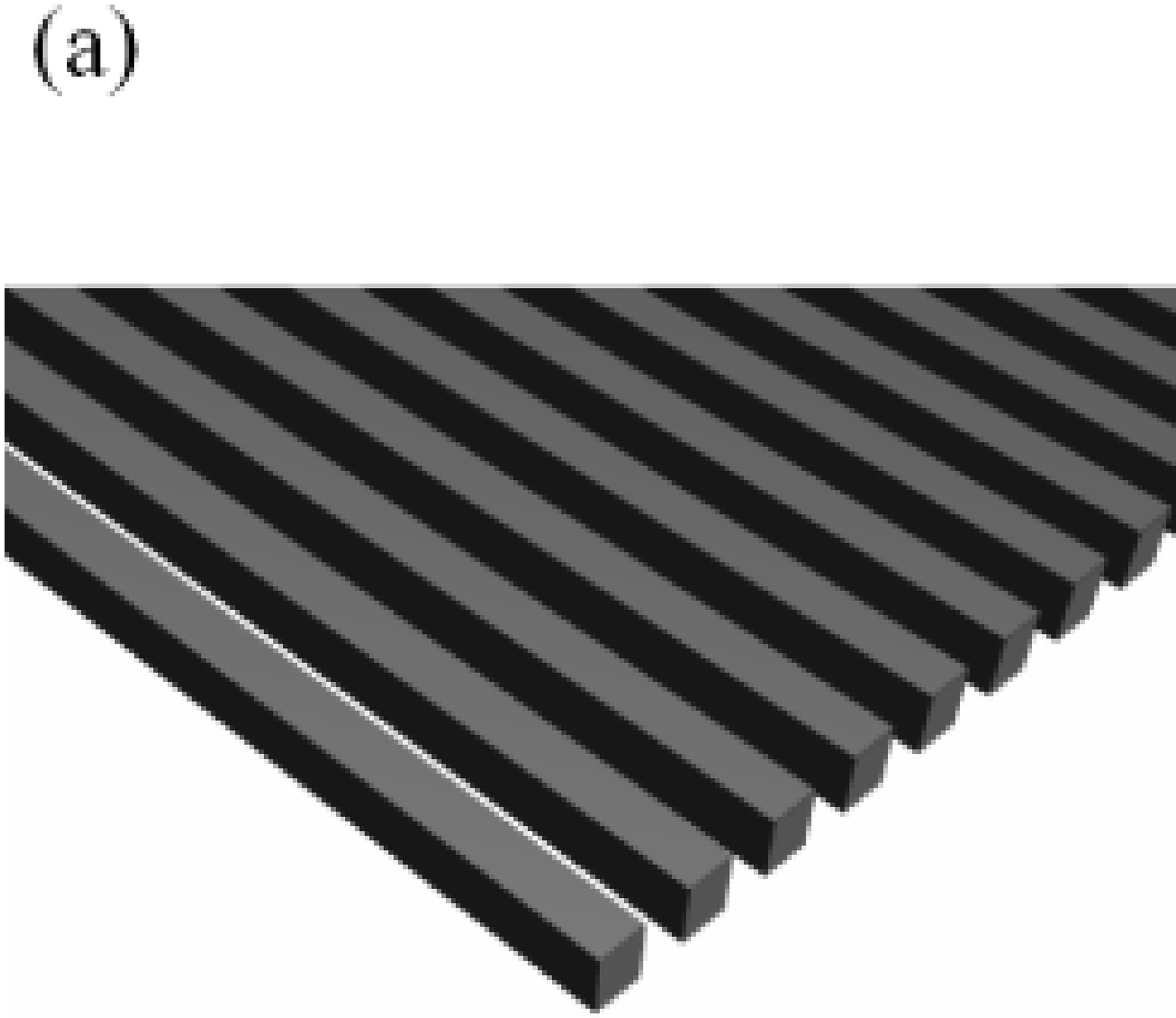}  &
\includegraphics[angle=90,width=3.6cm]{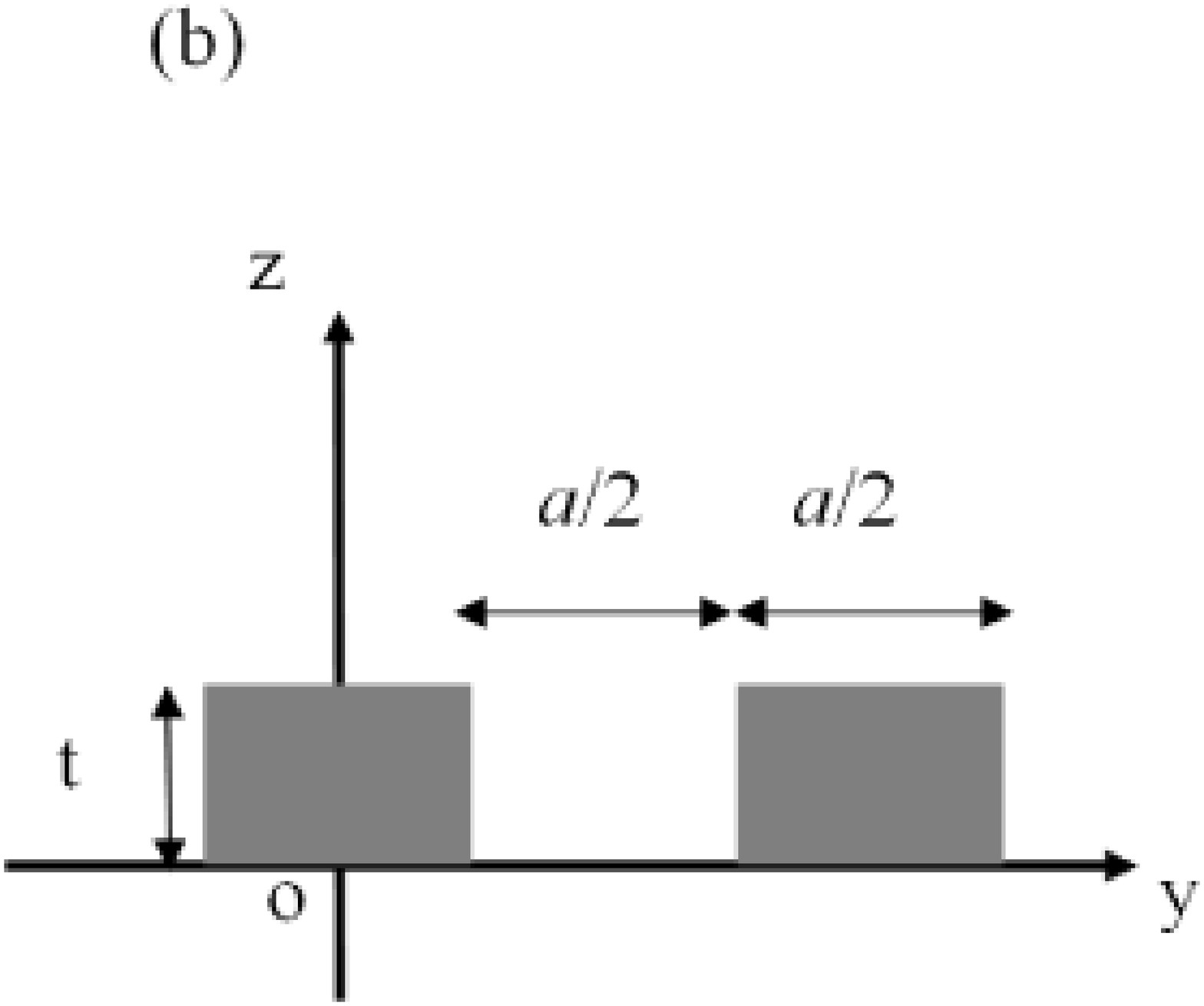} &
\includegraphics[angle=90,width=3.3cm]{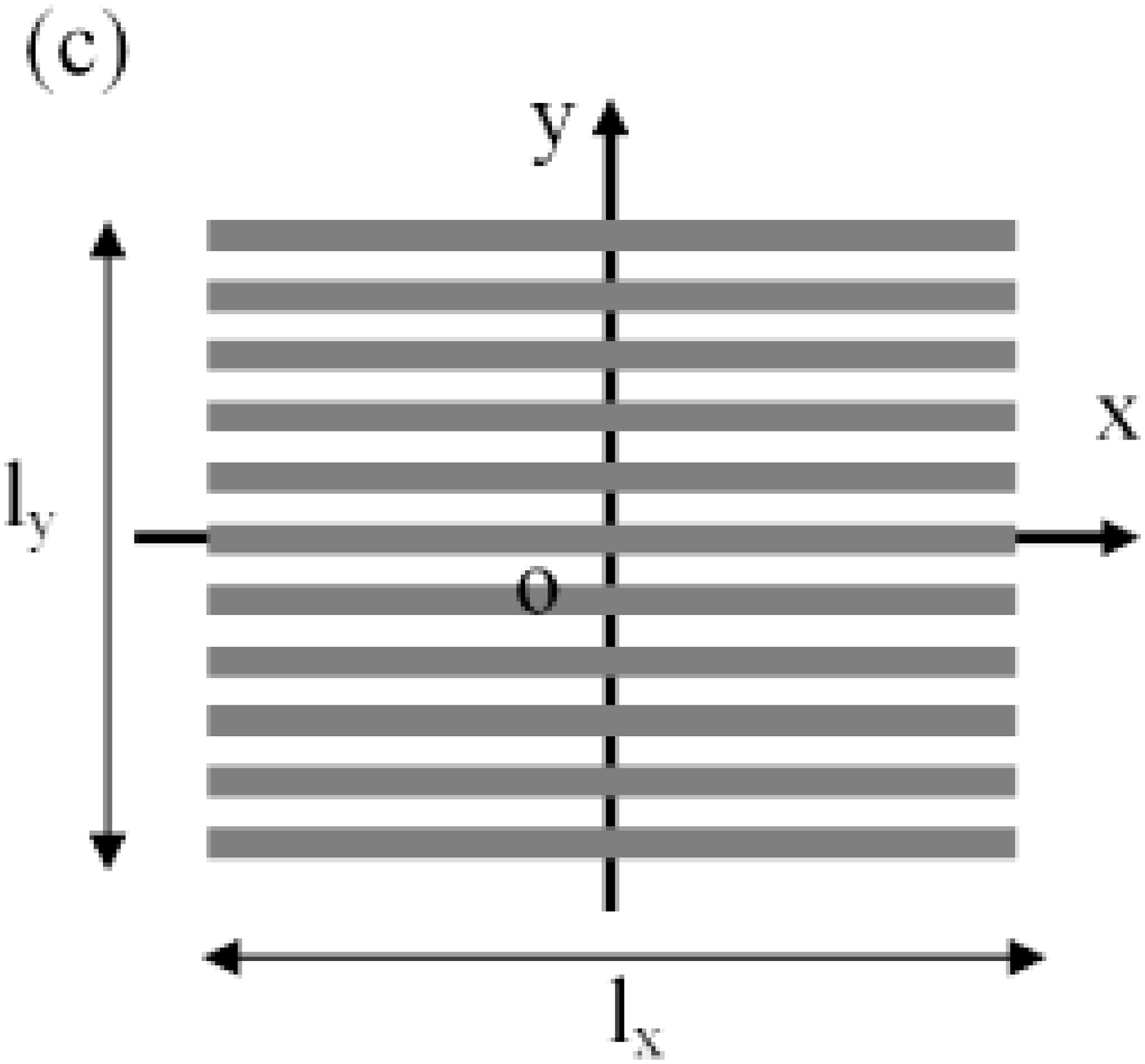}\\
\hspace{-.1cm}\includegraphics[angle=90,width=3.2cm,height=3.16cm]{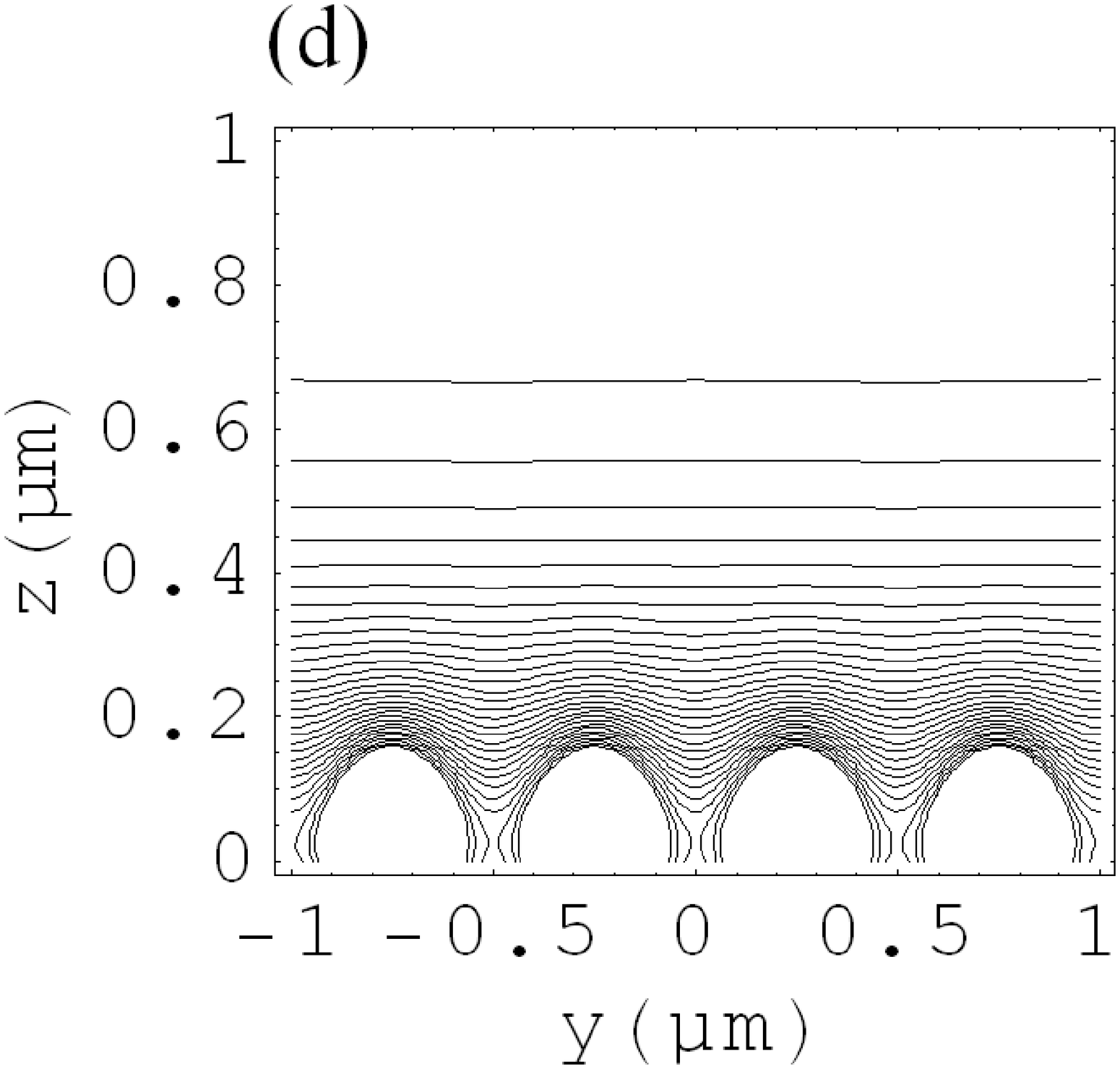}&
\hspace{-1.3cm}\includegraphics[angle=90,width=3.12cm,height=3.2cm]{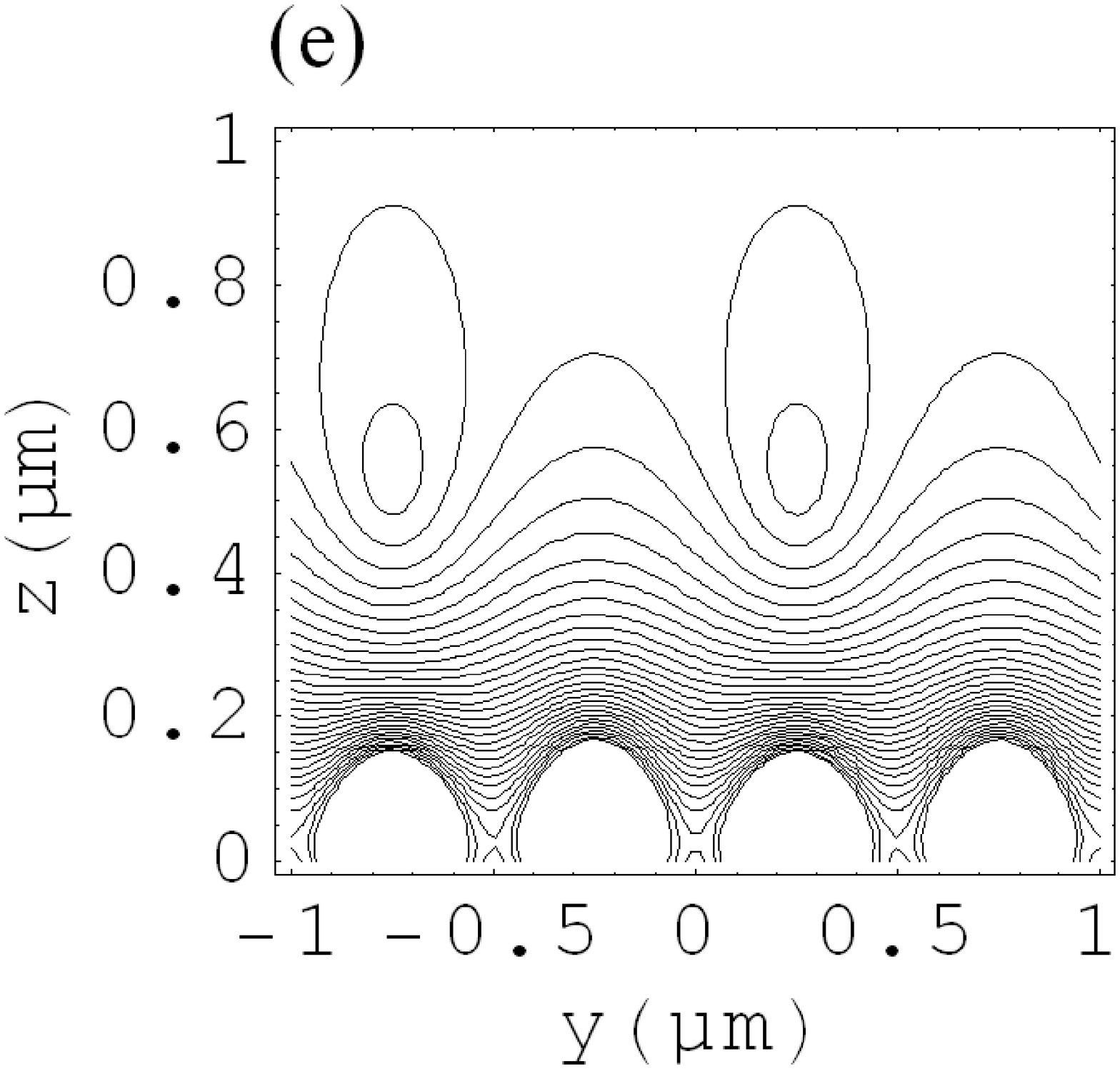}&
\hspace{-1.5cm}\includegraphics[angle=90,width=3.19cm,height=3.2cm]{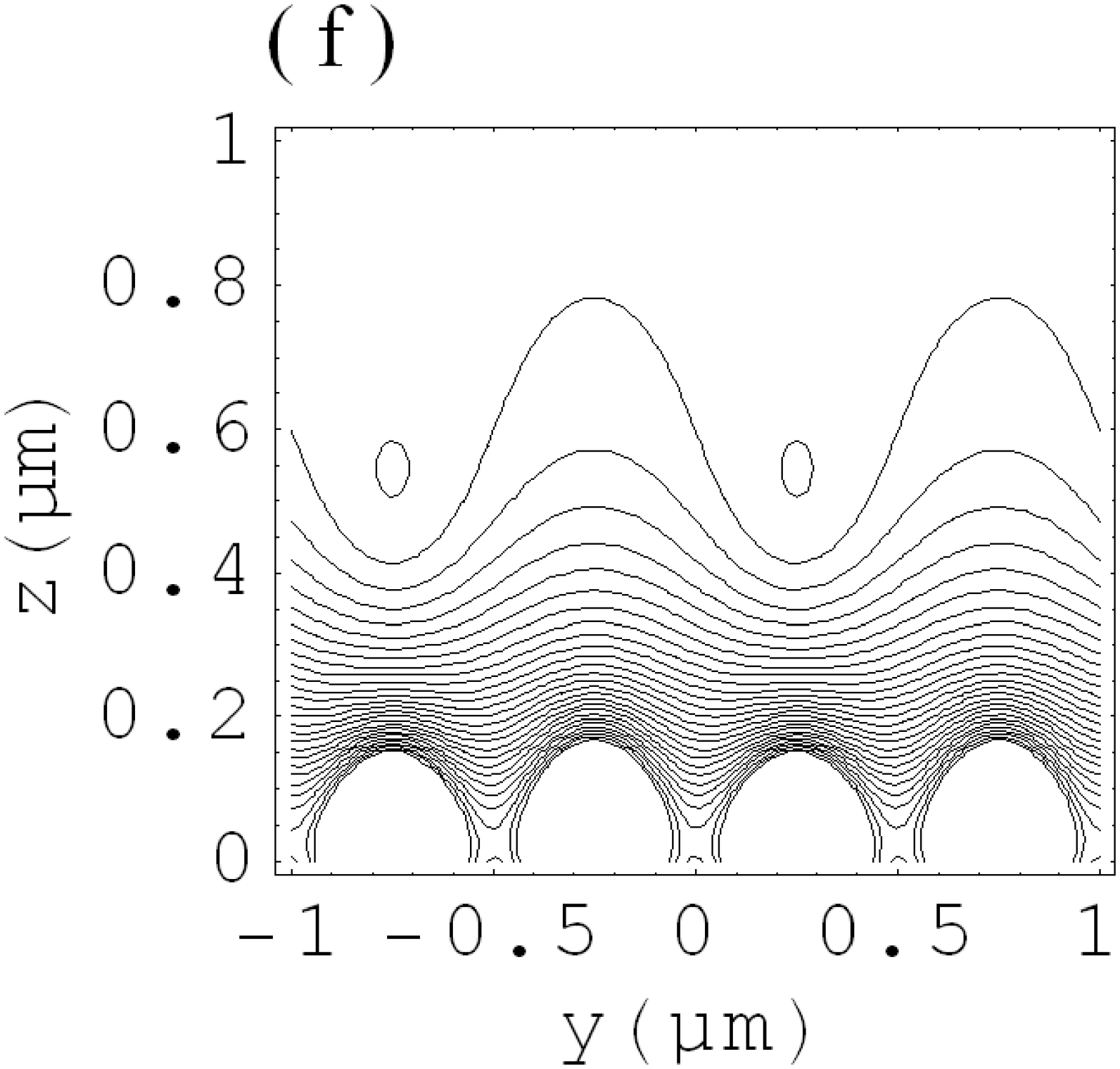}

\end{array}$
\caption{(a-c) Single periodic array of parallel, rectangular magnets with
perpendicular magnetization. (d-f) Contour plots of the magnitude of
the magnetic field in the central region in the $yoz$ plane without
bias fields (d), with a bias field $ B_{1y} =-15\hspace{.1cm} G $ along the
$y$-direction (e), and with bias fields $B_{1x} =- 20\hspace{.1cm} G$, $B_{1y} =-
15 \hspace{.1cm}G $ and $B_{1z} = -0.09 \hspace{.1cm}G$   along the $x$-, $y$- and
$z$-directions. For this calculation, the number of magnets $n_r = 1001$, $t = 0.05 \hspace{.1cm} \mu
m$, $a = 1 \hspace{.1cm}\mu m$, $l_x = 1000.5 \hspace{.1cm}\mu m$, and $4 \pi M_z = 3.8 \hspace{.1cm}kG$.
The spacing between the contour lines is $7\hspace{.1cm} G$. }
\label{figure1}
\end{center}
\end{figure}

We consider first the simple case of a single infinite periodic
array of parallel, rectangular, long magnets of thickness $t$, with
periodicity $a$ along the $y$-direction, perpendicular magnetisation
$M_z$, and uniform bias magnetic fields $B_{1x}$, $B_{1y}$ and
$B_{1z}$ along the $x$-, $y$- and $z$-directions [\fref{figure1}(a)-(c)].

The components of the magnetic field at distance $z$ from the bottom
surface of the array of magnets [\fref{figure1}(b)] can be written as (using the results given in~\cite{SidorovLau})
\numparts
\begin{eqnarray}\label{E1a}
 B_{x} = B_{1x}
\end{eqnarray}
\begin{eqnarray}\label{E1b}
 B_{y} &=& B_{0} [( 1-{\rm e}^{-kt}){\rm e}^{-k[z-(s+t)]}\sin(ky)\nonumber \\
&-& \frac{1}{3}( 1-{\rm e}^{-3kt}){\rm
e}^{-3k[z-(s+t)]}\sin(3ky)+\cdots] + B_{1y}
\end{eqnarray}
\begin{eqnarray}\label{E1c}
 B_{z} &=& B_{0} [( 1-{\rm e}^{-kt}){\rm e}^{-k[z-(s+t)]}\cos(ky)\nonumber \\
&-& \frac{1}{3}( 1-{\rm e}^{-3kt}){\rm
e}^{-3k[z-(s+t)]}\cos(3ky)+\cdots] + B_{1z}
\end{eqnarray}
\endnumparts
where the decay constant $k={2\pi/a}$, $B_0=4M_z$ (Gaussian units),
and the factors $(1 - e^{-kt})$, $(1 -  e^{-3kt})$, $\cdots$ account
for the finite thickness $t$ of the magnets.

For distances from the surface which are large compared with
$a/4\pi$, the higher order spatial harmonics in \eref{E1b} and
\eref{E1c} will have decayed away, and \eref{E1a}-\eref{E1c} reduce
to

\numparts
\begin{equation}\label{E2a}
B_{x} =  B_{1x}
\end{equation}
\begin{equation}\label{E2b}
B_{y} = B_{0y} \sin (ky) {\rm e}^{-kz}+ B_{1y}
\end{equation}
\begin{equation}\label{E2c}
B_{z} =  B_{0y} \cos (ky) {\rm e}^{-kz} + B_{1z}
\end{equation}
\endnumparts
where $ B_{0y} = B_{0} ( 1-{\rm e}^{-kt}){\rm e}^{kt}$.  The
magnitude of the magnetic field is then given by
\begin{eqnarray}\label{E3}
B(y,z) &=&\Bigl\{B_{1x}^2+B_{1y}^2+B_{1z}^2 \nonumber \\
&+& 2 [B_{0y}B_{1y} \sin(ky) + B_{0y}B_{1z} \cos(ky)] {\rm e}^{-kz}
+ B_{0y}^2 {\rm e}^{-2kz} \Bigr\}^{1\over 2}
\end{eqnarray}
The effect of the bias fields $B_{1y}$ and $B_{1z}$ is essentially the same; so we set $B_{1z}=0$.
This results in a 1D periodic
magnetic lattice of 2D magnetic traps with {\it non-zero} potential
minima, given by
\begin{equation}\label{E4}
B_{min} =  |B_{1x}|
\end{equation}
which are located at \numparts
\begin{equation}\label{E5a}
y_{min}=\left(n_y+{1 \over 4}\right) a
\end{equation}
\begin{equation}\label{E5b}
z_{min}={a \over {2\pi}} \ln \left({{B_{0y}}\over {|B_{1y}|}}\right)
\end{equation}
\endnumparts
where $n_y=0,\pm 1,\pm 2,\cdots$ represents the trap number in the
$y$-direction.  The barrier heights in the $y$-
and $z$-directions are given by
\numparts
\begin{equation}\label{E6a}
\Delta B^{y} =  B^y_{max}-B_{min}= (B_{1x}^2+ 4 B_{1y}^2)^{1\over 2}
-|B_{1x}|
\end{equation}
\begin{equation}\label{E6b}
\Delta B^{z} =  B^z_{max}-B_{min}= (B_{1x}^2+ B_{1y}^2)^{1\over 2}
-|B_{1x}|
\end{equation}
\endnumparts
The curvatures of the magnetic field at the centre of the 2D
magnetic traps and the trap frequencies  in the $y$- and $z$-directions are given by
\begin{equation}\label{E7}
\frac{\partial^2 B}{\partial y^2} =  \frac{\partial^2 B}{\partial z^2}= {{4\pi^2}\over a^2 } {B_{1y}^2\over |B_{1x}|}
\end{equation}
\begin{equation}\label{E08}
\omega_y= \omega_z = \frac{2\pi
}{a}\left(\frac{  m_{{}_F}
g_{{}_F}{\mu_{{}_B}}}{m |B_{1x}|}\right)^{\frac{1}{2}} |B_{1y}|
\end{equation}
where $m_{{}_F}$ is the magnetic quantum number of the hyperfine state
$F$, $g_{{}_F}$ is the Land\'{e} g-factor, $\mu_{{}_B}$ is the Bohr magneton and
$m$ is the atomic mass.
Equations \eref{E5b} and \eref{E08} are  the same as given in~\cite{SinclairCurtis}.

\begin{figure}[tbp]
\begin{center}
$\begin{array}{cc}
\includegraphics[angle=0,width=4.4cm,height=4.cm]{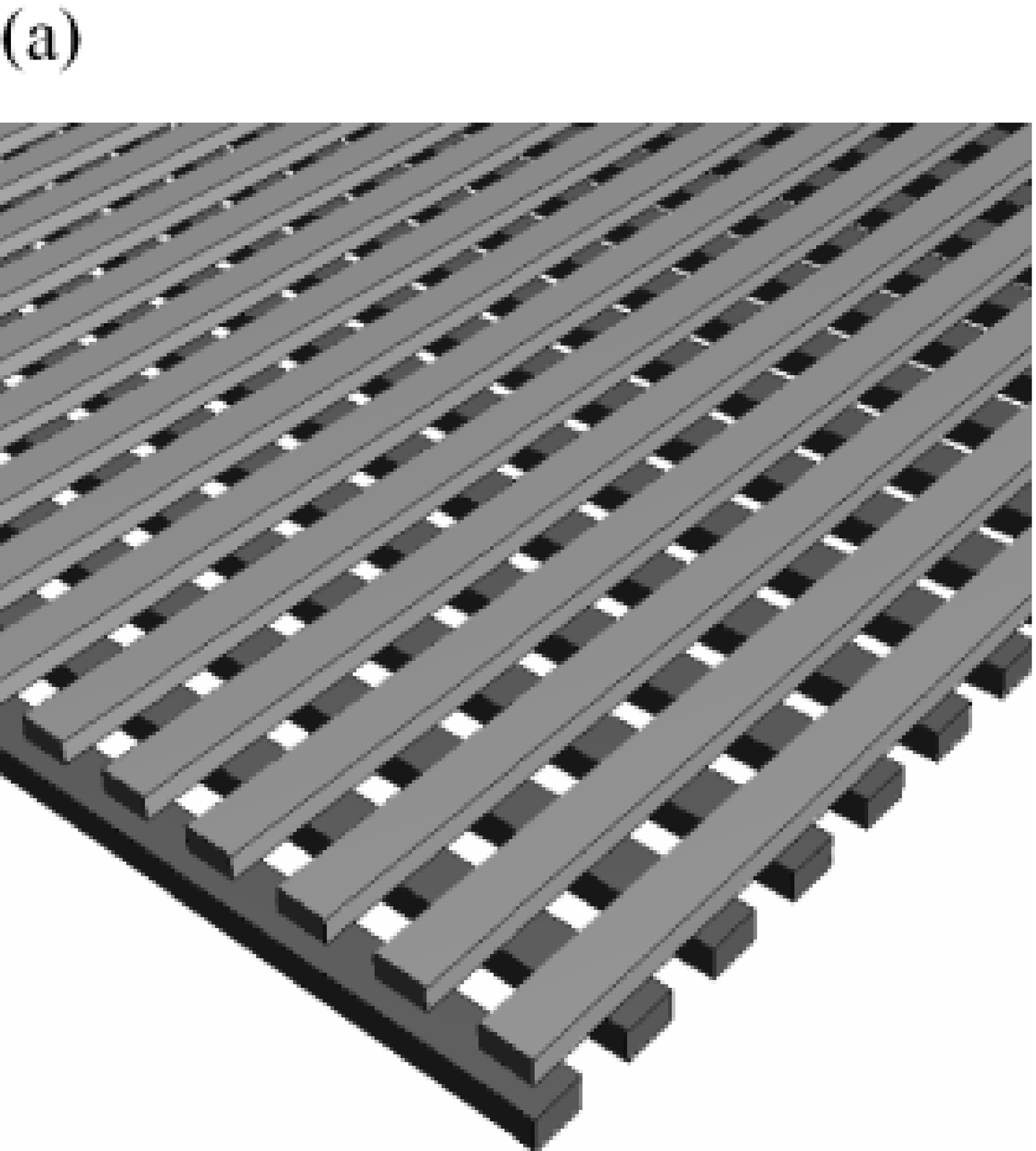}\label{fig2} &
\includegraphics[angle=0,width=4.4cm,height=4.cm]{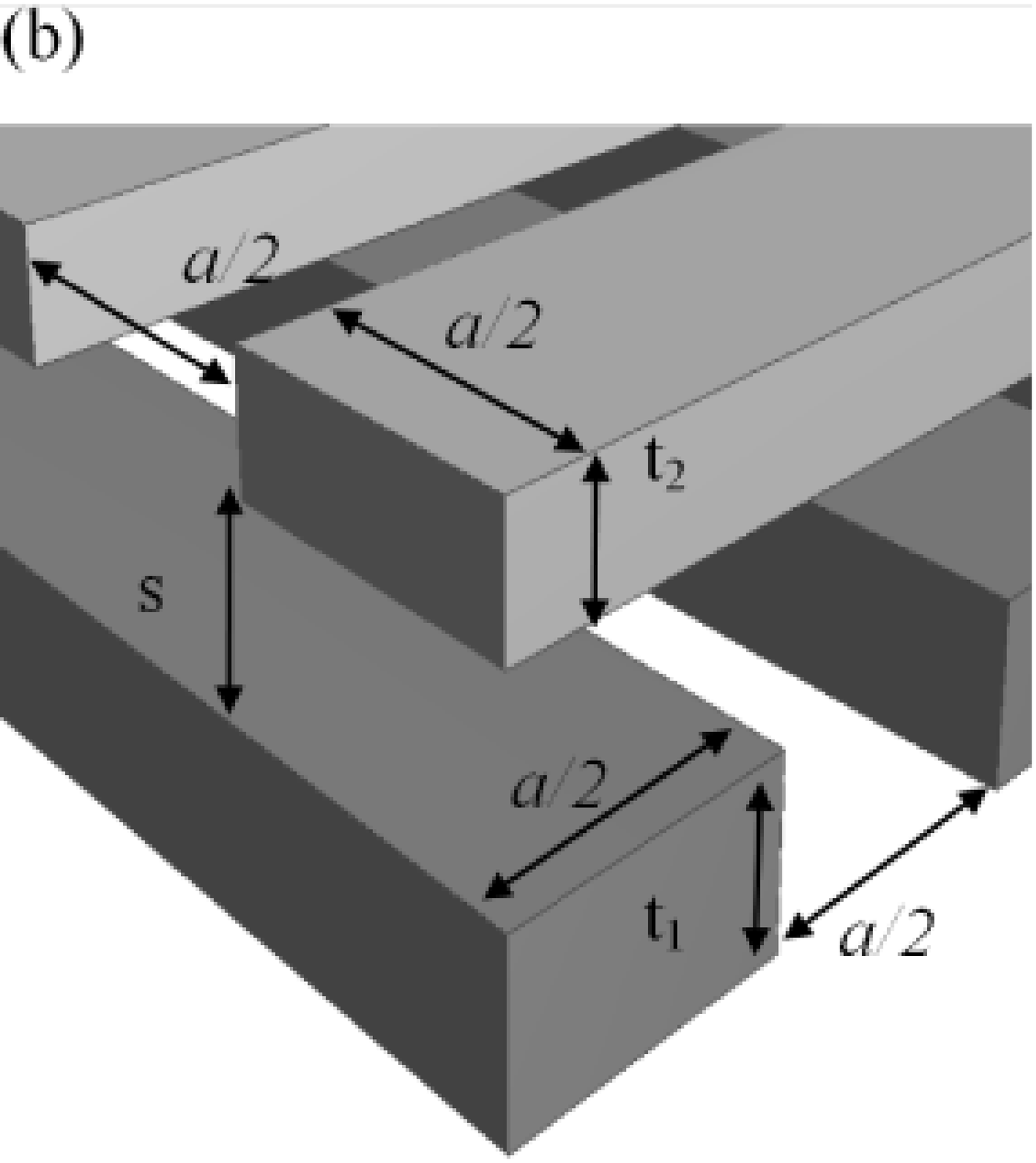}\\
\includegraphics[angle=90,width=4.8cm,height=4.cm]{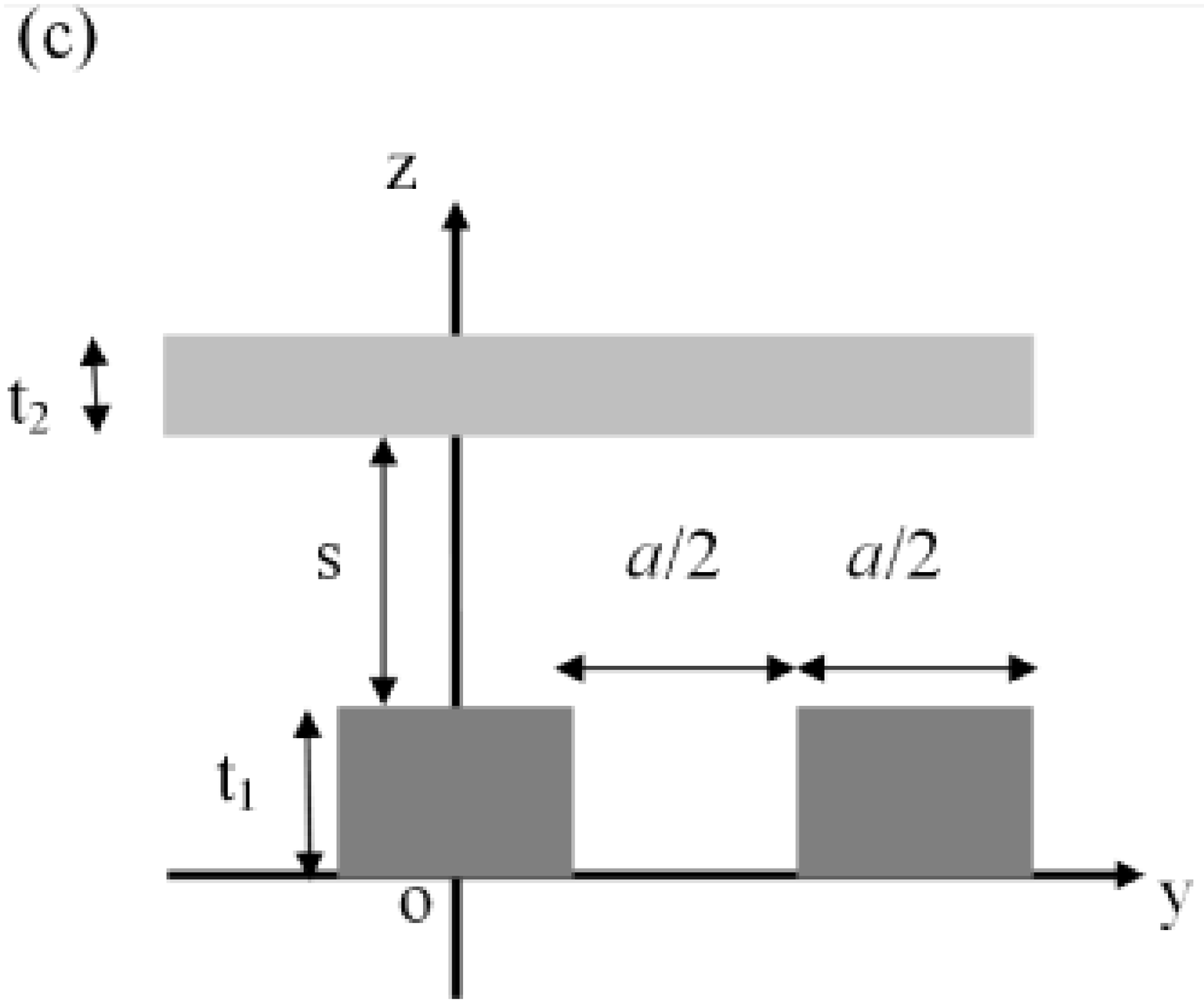}&
\includegraphics[angle=0,width=4.4cm,height=4.cm]{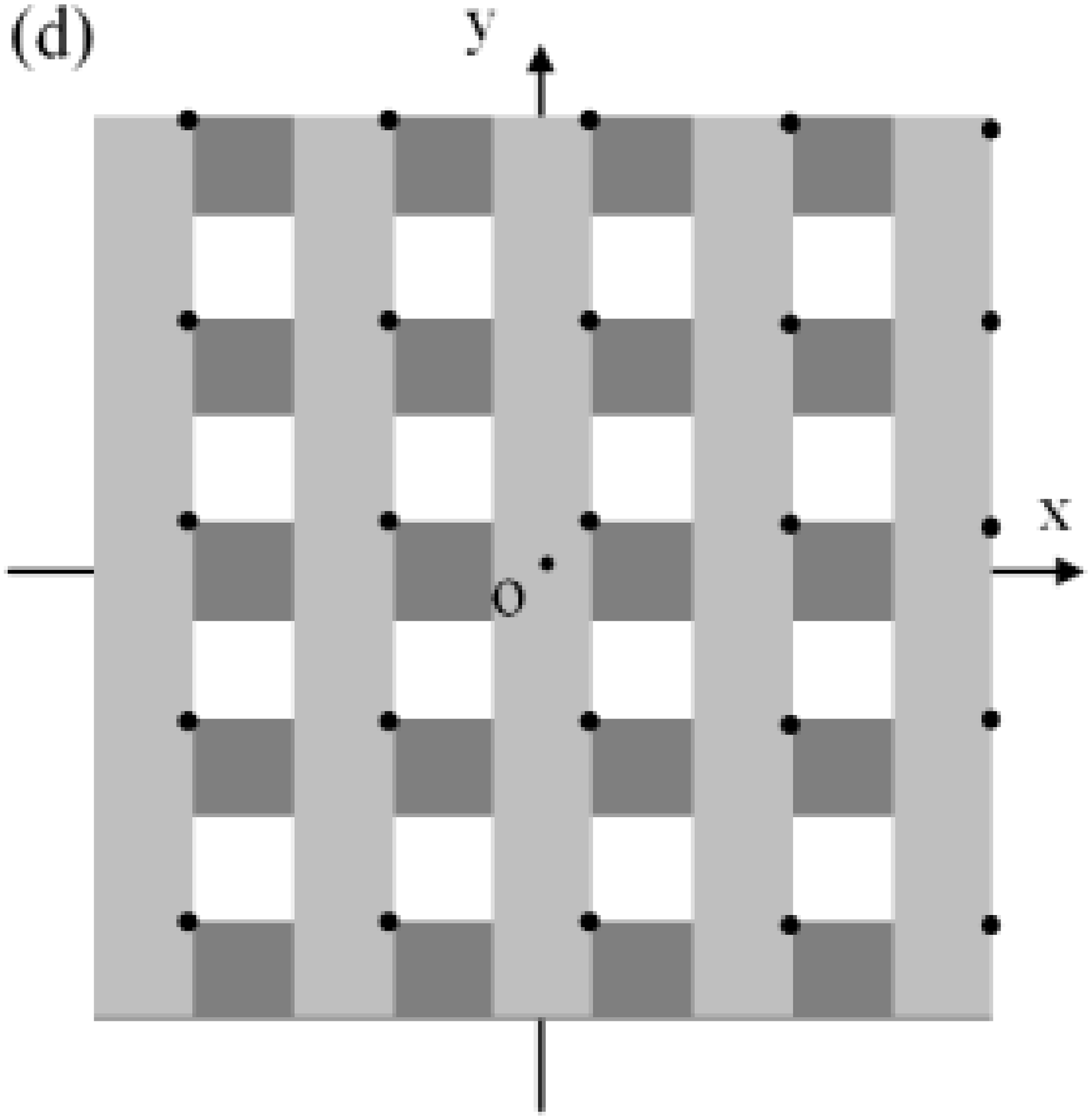}
\end{array}$
\caption{Periodic array consisting of two crossed layers of
parallel, rectangular permanent magnets with perpendicular
magnetisation.  In (d) the locations of the central minima in the
$xoy$ plane are shown for a symmetrical magnetic lattice with bias
fields $B_{1x}= - 4.08 \hspace{.1cm}G$, $B_{1y} = - 6.05 \hspace{.1cm}G$ and $B_{1z} = - 0.69\hspace{.1cm}
G$ and other parameters as given in table 1, column 3. }
\label{figure2}
\end{center}
\end{figure}

\subsubsection{No bias fields}
 For the case of no bias fields, \eref{E3} reduces to
\begin{equation}\label{E8}
B(z)= B_{0y}{\rm e}^{-kz}
\end{equation}
Under these conditions the magnitude of the magnetic field falls off
exponentially with distance z from the surface.  This configuration
represents the familiar case of a grooved magnetic mirror for slowly
moving atoms in low magnetic field-seeking
states~\cite{SidorovMcSch}.

\subsubsection{Single bias field $B_{1y}$}

For the case of a single bias field $B_{1y}$, \eref{E3} becomes
\begin{equation}\label{E9}
B(y,z) = \Bigl\{B_{1y}^2  + 2 B_{0y}B_{1y}\sin(ky)  {\rm e}^{-kz} +
B_{0y}^2 {\rm e}^{-2kz} \Bigr\}^{1\over 2}
\end{equation}
The magnitude of the magnetic field develops corrugations with
period $a$ in the $y$-direction, and 2D magnetic traps with $B_{min} =
0$ appear in the potentials above the surface.  This configuration, involving a single bias field,
may be used as a spatial diffraction grating for slowly moving
atoms~\cite{OpatWark,Lau,OpatChormaic,Rosenbusch}, but is not useful as a
magnetic lattice for ultracold atoms because of the zero potential minima.

\subsection{Two crossed layers of infinite periodic arrays of magnets with bias
fields}\label{sec2.2} We now consider a configuration consisting of
two crossed, separated, infinite periodic arrays of parallel,
rectangular, long magnets with perpendicular magnetization $M_z$ and
uniform bias fields $B_{1x}$, $B_{1y}$ and $B_{1z}$ along the $x$-,
$y$- and $z$-directions (\fref{figure2}).
\begin{table}[hb]
 \caption{\label{table1}Input parameters for (1) two crossed
layers of periodic arrays of rectangular magnets (\fref{figure2}),
and (2) a single layer of square magnets with three thicknesses
(\fref{figure7}).}
\begin{indented}
\item[]
\hspace{-1.2cm}
\begin{tabular}{@{}llll}
\br
Parameter&Definition&Configuration (1)&Configuration (2)\\
\mr
\verb   $n_r$ or $n_{sq}$    & Number of rectangular          &  $n_r = 1001$                &   $n_{sq} = 401$                 \\
\verb                        & magnets or square magnets      &                              &                                  \\
\verb                        & in $x$- or $y$-direction       &                              &                                  \\
\verb   $a\hspace{.1cm}(\mu m)$           & Period of magnetic lattice     &  $1.000 $                   &   $1.000 $                      \\
\verb   $l_x=l_y\hspace{.1cm}(\mu m)$            &  Length of magnets  or         &  $1000.5 $              &   $200.5 $                  \\
\verb                        & magnetic array along $x$ or $y$&                              &                                  \\
\verb   $s\hspace{.1cm}(\mu m)$                  & Separation of magnetic layers  &  $0.100 $               &                                  \\
\verb   $t_1\hspace{.1cm}(\mu m)$                &  Thickness of magnetic         &  $0.322 $               &   $0.120 $                  \\
\verb                        &  film (first)                  &                              &                                  \\
\verb   $t_2\hspace{.1cm}(\mu m)$                & Thickness of magnetic          &  $0.083 $               &   $0.100 $                  \\
\verb                        &  film (second)                 &                              &                                  \\
\verb   $t_3\hspace{.1cm}(\mu m)$                &  Thickness of  magnetic        &                              &   $0.220 $                  \\
\verb                        & film (third)                   &                              &                                  \\
\verb   $4 \pi M_z\hspace{.1cm}(G)$          & Magnetisation along z          &  $3800 $                    &   $3800 $                       \\
\verb   $B_{1x}\hspace{.1cm}(G)$             & Bias magnetic field along $x$ &  $\hspace{-.33cm}$ $-4.08 $  &   $\hspace{-.33cm}$  $-5.00 $    \\
\verb   $B_{1y}\hspace{.1cm}(G)$             & Bias magnetic field along $y$  &  $\hspace{-.33cm}$ $-6.05 $  &   $\hspace{-.33cm}$ $-4.22 $     \\
\verb   $B_{1z}\hspace{.1cm}(G)$             & Bias magnetic field along $z$  &  $\hspace{-.33cm}$ $-0.69 $  &   $\hspace{-.33cm}$ $-1.87 $     \\
\br
\end{tabular}
\end{indented}
\end{table}
The bottom array has periodicity $a$
along the $y$-direction and the top array has  periodicity $a$
along the $x$-direction. The two arrays are separated by a distance
$s$ and the magnets in the bottom array have thickness $t_1$ while
those in the top array have thickness $t_2$ [see \fref{figure2}(b)].
This configuration of crossed permanent magnets with bias
fields has similarities to the crossed current-carrying wire
configurations proposed by Yin \etal~\cite{Yin} and by Grabowski and
Pfau~\cite{Grabowski}.
The components of the magnetic field for distances from the top
surface which are large compared with $a/{4\pi}$  are given by

\numparts
\begin{equation}\label{E10a}
B_{x} = B_{0x} \sin (kx) {\rm e}^{-kz}+ B_{1x}
\end{equation}
\begin{equation}\label{E10b}
B_{y} = B_{0y} \sin (ky) {\rm e}^{-kz}+ B_{1y}
\end{equation}
\begin{equation}\label{E10c}
B_{z} = [ B_{0x} \cos (kx)+B_{0y} \cos (ky)] {\rm e}^{-kz}+B_{1z}
\end{equation}
where
\endnumparts
\numparts
\begin{equation}\label{E11a}
B_{0x} =B_{0} (1- {\rm e}^{-kt_2}){\rm e}^{k(s+t_1+t_2)}
\end{equation}
\begin{equation}\label{E11b}
B_{0y} = B_{0} ( 1-{\rm e}^{-kt_1}){\rm e}^{kt_1}
\end{equation}
\endnumparts

The magnitude of the magnetic field above the crossed magnetic
arrays is then
\begin{equation*}
 B(x,y,z) = \Bigl\{B_{1x}^2+B_{1y}^2 +B_{1z}^2  \nonumber
\end{equation*}
\begin{equation*}
 \qquad  \qquad + 2 [B_{0x}B_{1x} \sin(kx) + B_{0y}B_{1y} \sin(ky)
\end{equation*}
\begin{equation*}
 \qquad  \qquad + B_{0x}B_{1z} \cos(kx) + B_{0y}B_{1z} \cos(ky)] {\rm e}^{-kz}
\nonumber\
\end{equation*}
\begin{equation}
  \qquad  \qquad +[B_{0x}^2+B_{0y}^2 + 2 B_{0x}B_{0y} \cos(kx)\cos(ky) ]{\rm
e}^{-2kz} \Bigr\}^{1\over 2} \label{E12}
\end{equation}

\subsubsection{Two bias fields $B_{1x}$ and $B_{1y}$}\label{2.2.1}
For the case of two bias fields $B_{1x}$ and $B_{1y}$, \eref{E12}
becomes

\begin{equation*}
 B(x,y,z) = \Bigl\{B_{1x}^2+B_{1y}^2
\end{equation*}
\begin{equation*}
 \qquad  \qquad + 2 [B_{0x}B_{1x} \sin(kx) + B_{0y}B_{1y} \sin(ky)]{\rm e}^{-kz}
\end{equation*}
\begin{equation}
 \qquad  \qquad +[B_{0x}^2+B_{0y}^2 + 2 B_{0x}B_{0y} \cos(kx)\cos(ky) ]{\rm
e}^{-2kz} \Bigr\}^{1\over 2} \label{E13}
\end{equation}
This configuration results in a 2D periodic lattice of magnetic
traps with {\it non-zero} potential minima given by
\begin{equation}\label{E14}
B_{min}=  \frac{|B_{0x}B_{1y}  -
B_{0y}B_{1x}| }{({B_{0x}^2+B_{0y}^2})^{1\over 2}}
\end{equation}

\subsubsection{Symmetrical magnetic lattice with
bias fields $B_{1x}$ and $B_{1y}$}

To produce a symmetrical magnetic lattice the amplitude of the
oscillating magnetic field produced by the bottom magnetic array in
the $y$-direction needs to equal the amplitude of the oscillating
magnetic field produced by the top array in the $x$-direction.  To
satisfy this condition we impose the constraint
\numparts
\begin{equation}\label{E15a}
B_{0x}B_{1x} = B_{0y}B_{1y}
\end{equation}
or
\begin{equation}\label{E15b}
B_{1y} =c_0 B_{1x}
\end{equation}
\endnumparts
where
\begin{equation}\label{E16}
c_0 = \frac{B_{0x}}{B_{0y}} = \left(\frac{{\rm e}^{kt_2} - 1}{1 - {\rm
e}^{-kt_1}}\right){\rm e}^{ks}
\end{equation}
is a dimensionless constant which only involves geometrical constants $a, s, t_1$ and $t_2$ of the magnetic array.
The magnetic traps then have  non-zero potential minima given
by
\begin{equation}\label{E17}
B_{min}=c_1 |B_{1x}|
\end{equation}
which are located at
\numparts
\begin{equation}\label{E18a}
x_{min}=\left(n_x+{1 \over 4}\right) a,  \qquad n_x=0,\pm 1,\pm
2,\cdots
\end{equation}
\begin{equation}\label{E18b}
y_{min}=\left(n_y+{1 \over 4}\right) a,  \qquad  n_y=0,\pm 1,\pm
2,\cdots
\end{equation}
\begin{equation}\label{E18c}
z_{min}={a \over {2\pi}} \ln \left({{c_2 B_{0x}}\over
|B_{1x}|}\right)
\end{equation}
\endnumparts
where $c_1$ and $c_2$ are dimensionless constants which may be
expressed in terms of the constant $ c_0=B_{0x}/B_{0y}$
\numparts
\begin{equation}\label{E19a}
c_1= \frac{|1 - c_0^2|}{{(1 +c_0^2)}^{1\over 2}}
\end{equation}
\begin{equation}\label{E19b}
c_2 = {1\over 2}\left(1 + \frac{1}{c_0^2}\right)
\end{equation}
\endnumparts
In order to have non-zero potential minima above the surface of the
top array, we have the following constraints
\begin{equation}\label{E20}
c_2 B_{0x}> |B_{1x}| > 0 , \qquad  c_0 c_2 B_{0y} > |B_{1y}|
> 0, \qquad  B_{0x} \ne B_{0y}
\end{equation}

The curvatures of the magnetic field at the centre of the traps and
the trap frequencies (for the case of a harmonic potential) in the
three directions are given by
\begin{equation}\label{E21}
\frac{\partial^2 B}{\partial x^2}=\frac{\partial^2 B}{\partial y^2}={1\over 2}\frac{\partial^2 B}{\partial z^2} = \frac{4 \pi^2 c_3}{a^2} |B_{1x}|
\end{equation}
\begin{equation}\label{E22}
\omega_x= \omega_y={\omega_z \over \sqrt{2}} = \frac{2\pi
}{a}\left(\frac{  m_{{}_F}
g_{{}_F}{\mu_{{}_B}c_3}}{m}\right)^{\frac{1}{2}} |B_{1x}|^{1\over 2}
\end{equation}
The potential barrier heights in the three
directions are given by
\numparts
\begin{equation}\label{E23a}
\Delta B^x = \Delta B^y =c_4 |B_{1x}|
\end{equation}
\begin{equation}\label{E23b}
\Delta B^z=c_5|B_{1x}|
\end{equation}
\endnumparts
$c_3$, $c_4$ and $c_5$ are dimensionless constants which may be
expressed in terms of $c_0$
\numparts
\begin{equation}\label{E24a}
c_3=\frac{2 c_0^2}{(1+c_{0}^2)^{\frac{1}{2}}|1-c_{0}^2|}
\end{equation}
\begin{equation}\label{E24b}
c_4= \left(1 + c_0^2 + \frac{4 c_0^2 }{1 + c_0^2}\right)^{1\over 2}
- \frac{|1 - c_0^2|}{(1 + c_0^2)^{1\over 2}}
\end{equation}
\begin{equation}\label{E24c}
c_5=(1+c_{0}^2)^{1\over 2}
\end{equation}
\endnumparts
The barrier heights in the $x$-, $y$- and $z$-directions vary with
$z_{min}$ as
\numparts
\begin{equation}\label{E25a}
\Delta B^x = \Delta B^x_0 {\rm e}^{-k(z_{min}-z_{min0})} = \Delta
B^y = \Delta B^y_0 {\rm e}^{-k(z_{min}-z_{min0})}\hspace{.5cm}
\end{equation}
\begin{equation}\label{E25b}
\Delta B^z=\Delta B^z_0 {\rm e}^{-k(z_{min}-z_{min0})}
\end{equation}
\endnumparts
where \numparts
\begin{equation}\label{E26a}
\Delta B^x_0 = \Delta B^y_0 =c_1 c_4 B_{0x}{\rm e}^{-kz_{min0}}
\end{equation}
\begin{equation}\label{E26b}
\Delta B^z_0 =c_1 c_5 B_{0x}{\rm e}^{-kz_{min0}}
\end{equation}
\endnumparts
If $B_{1x}$ and $B_{1y}$  are not subject to the constraint
\eref{E15a} we have $\Delta B^x \ne \Delta B^y$. In some
experiments, e.g., in studies of quantum tunnelling between lattice
sites, it may be useful to be able to have different barrier heights
along  different axes.
\subsubsection{Three bias fields $B_{1x}$, $B_{1y}$ and $B_{1z}$}\label{sec2.2.3}
For the case of three bias fields $B_{1x}$, $B_{1y}$ and $B_{1z}$,
the magnitude of the magnetic field is given by \eref{E12}.  The
analytical expressions for $B_{min}$, $x_{min}$, $y_{min}$ and
$z_{min}$ in this case become very complex and we resort to
numerical evaluation of \eref{E12} to determine these quantities.
Moreover, the condition for a symmetrical magnetic lattice in the
case $B_z\ne 0$ imposes the constraints
\begin{equation}\label{E27}
B_{0x} = B_{0y}, \qquad B_{1x} = B_{1y}
\end{equation}
and under such conditions, $B_{min} = 0$.  Thus in the case of an {\it
infinite} symmetrical magnetic lattice with three bias fields,
$B_{1x}$, $B_{1y}$ and $B_{1z}$, it is not possible to have non-zero
potential minima.

Nevertheless, we find that in the case of a {\it finite} magnetic lattice
it is useful  to be able to apply a bias field $B_{1z}$ in order
to compensate for the asymmetry introduced into the lattice by
end-effects~\cite{SidorovMcRow} associated with the finite number of
magnets in the array  (see \sref{sec3}).

\section{Numerical calculations for finite magnetic lattices}\label{sec3}
 To calculate magnetostatic potentials for arbitrary configurations,
including finite periodic arrays of magnets of finite length and
arbitrary cross section, the software package Radia~\cite{esrf}
interfaced to Mathematica was used.  Radia was also used to test
various proposed configurations of permanent magnets that might
support periodic arrays of microtraps with non-zero potential minima
based on symmetry arguments.

\subsection{Single finite periodic array of
magnets with bias fields}\label{sec3.1} \Fref{figure1}(d) shows a
numerical calculation of the magnetic field versus distance in the
$y$-direction for the central region of a single finite array of
$n_r=1001$ rectangular magnets, using the parameters $a = 1 \hspace{.1cm} \mu m$,
$t = 0.050 \hspace{.1cm}\mu m$, $l_x = 1000.5 \hspace{.1cm}\mu m$ and $4 \pi Mz = 3.8 \hspace{.1cm}kG$
(corresponding to the magnetization of our perpendicularly
magnetized $Tb_6Gd_{10}Fe_{80}Co_4$ magneto-optical
films~\cite{Wang,Hall}) and with {\it no bias} magnetic fields.
Corrugations corresponding to the third-order spatial harmonic with
period $a/2$~\cite{SidorovLau} [see \eref{E1b} and \eref{E1c}] can
be seen at distances very close ($\ll {a}/{4 \pi}$) to the
surface. At large distances, there are small residual corrugations
with period $a$, which are due to `end effects'~\cite{SidorovMcRow}
associated with the finite number ($1001$) of magnets in the array.
This configuration may be used as a magnetic mirror for slowly
moving atoms~\cite{SidorovMcSch}.

\Fref{figure1}(e) shows the effect of adding a bias field $B_{1y} =-
15 \hspace{.1cm}G$. The magnitude of the magnetic field develops large
corrugations with period $a$ in the $y$-direction, and 2D magnetic
traps with $B_{min} = 0$ appear in the potentials at $z_{min} =
0.540 \hspace{.1cm}\mu m$.  This configuration may be used as a spatial
diffraction grating~\cite{OpatWark,Lau,OpatChormaic,Rosenbusch} for slowly
moving atoms.

\Fref{figure1}(f) shows the effect of having bias fields $B_{1x} =-
20 \hspace{.1cm}G$, $B_{1y} =-15 \hspace{.1cm}G$ and $B_{1z} =- 0.09 \hspace{.1cm}G$.  The small value of
$B_{1z}$ was chosen in order to compensate for asymmetry introduced into
the lattice by end-effects~\cite{SidorovMcRow} associated with the
finite number of magnets (1001) in the array.
\begin{figure}[tbp]
\begin{center}
$\begin{array}{ccc}
\hspace{.3cm}\includegraphics[angle=90,width=3.5cm,height=3.2cm]{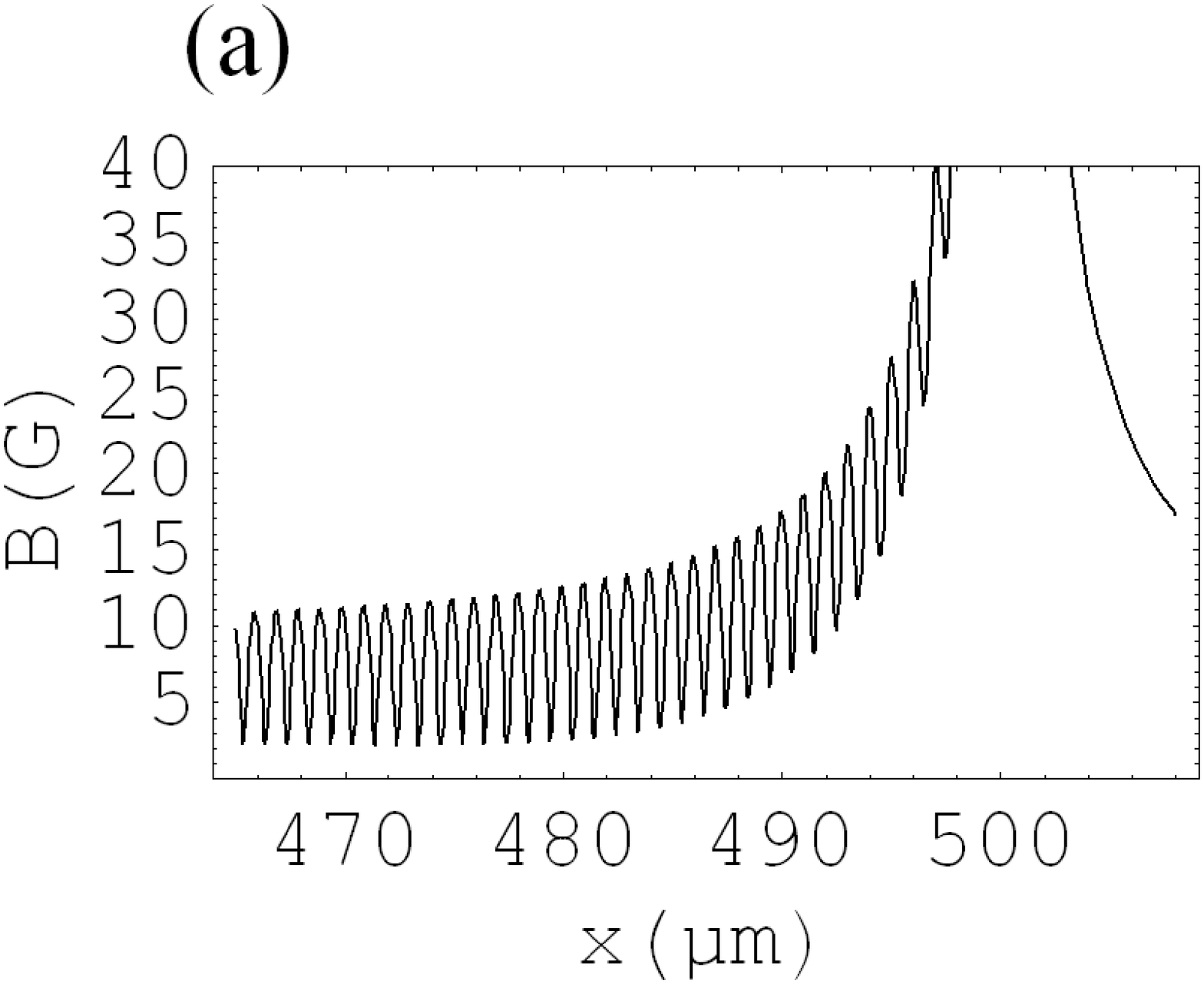} &
\hspace{-.0cm}\includegraphics[angle=90,width=3.5cm,height=3.2cm]{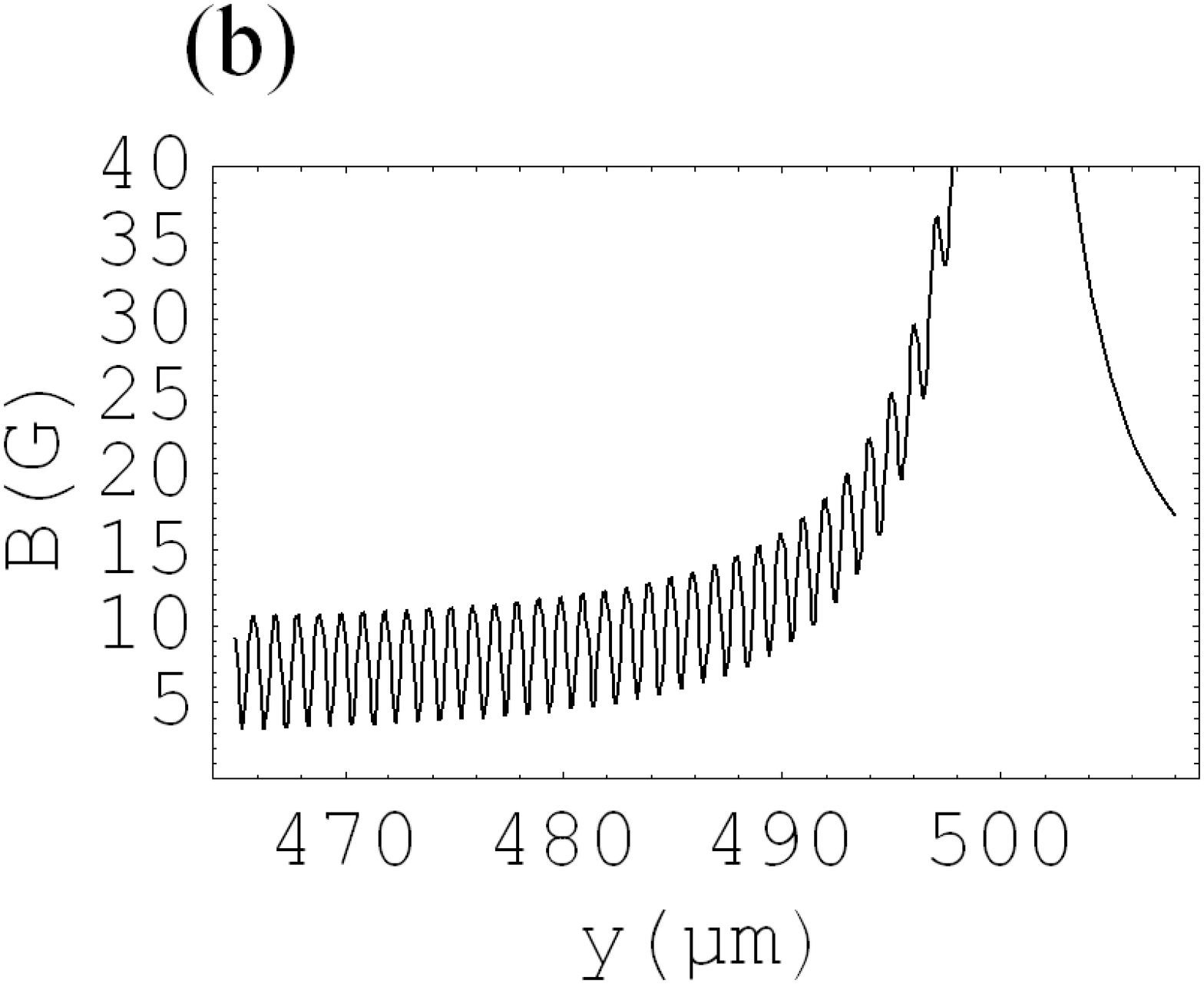} &
\hspace{-.6cm}\includegraphics[angle=90,width=3.5cm,height=3.3cm]{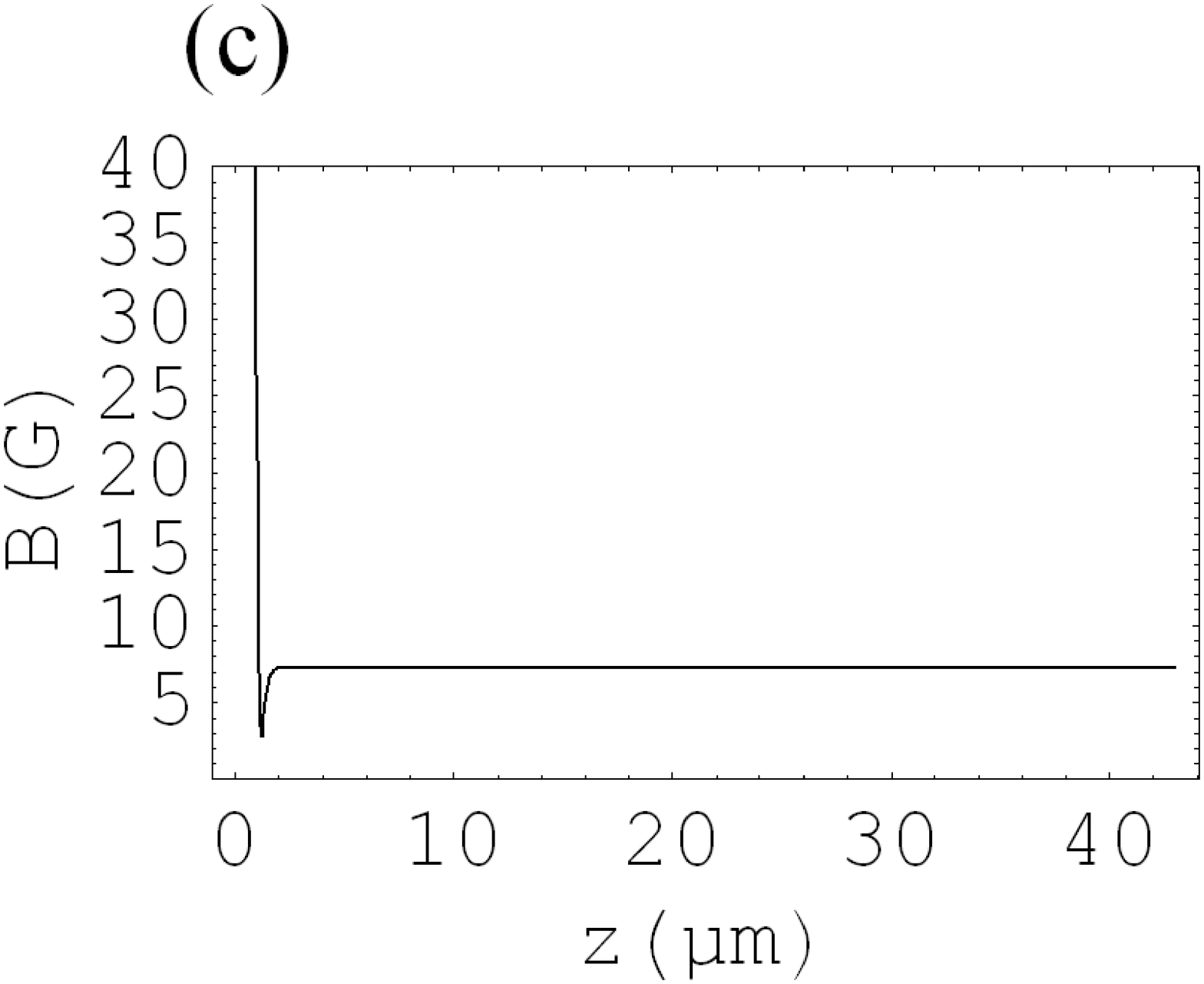}\\
\hspace{.2cm}\includegraphics[angle=90,width=3.62cm,height=3.2cm]{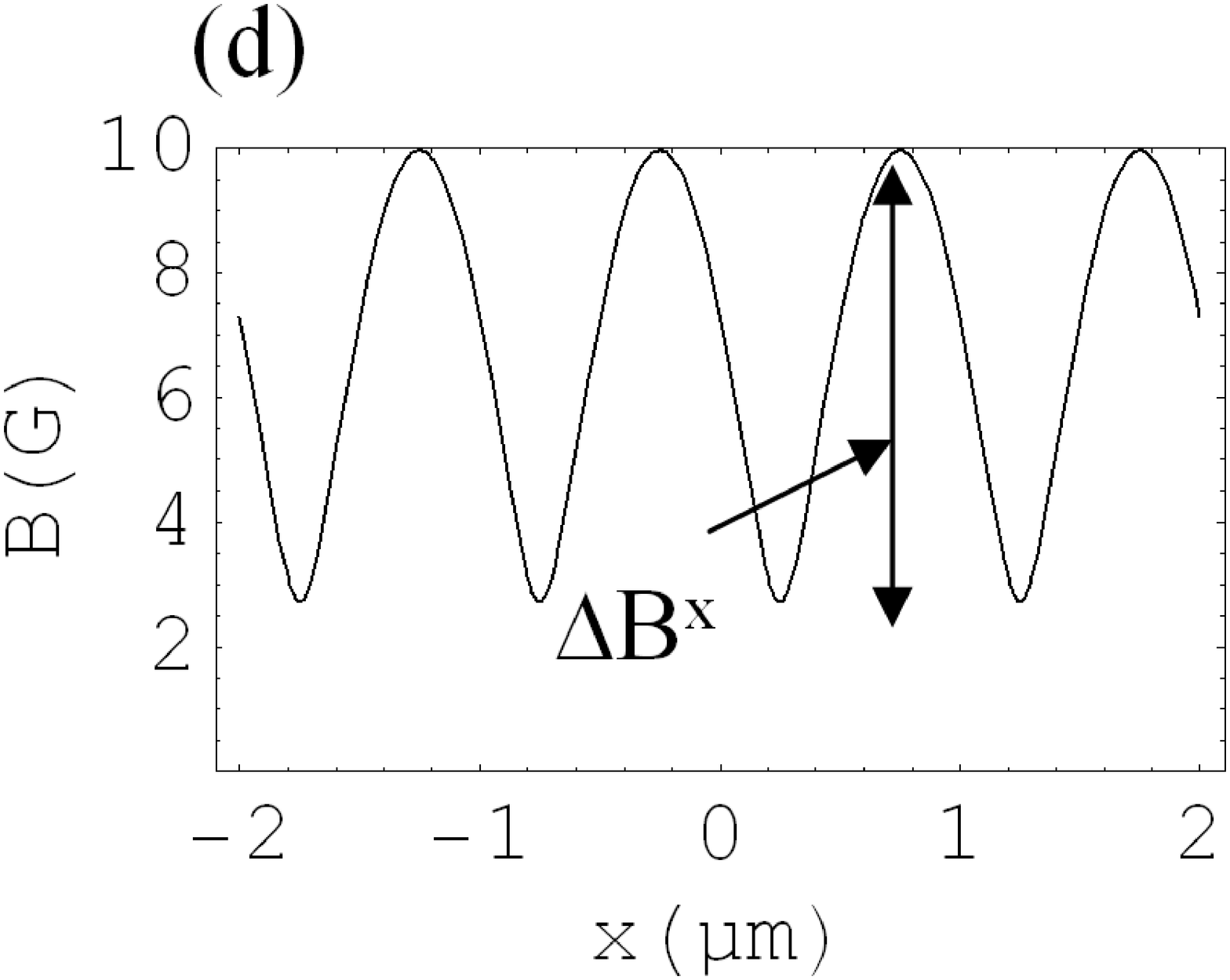} &
\hspace{.2cm}\includegraphics[angle=90,width=3.62cm,height=3.2cm]{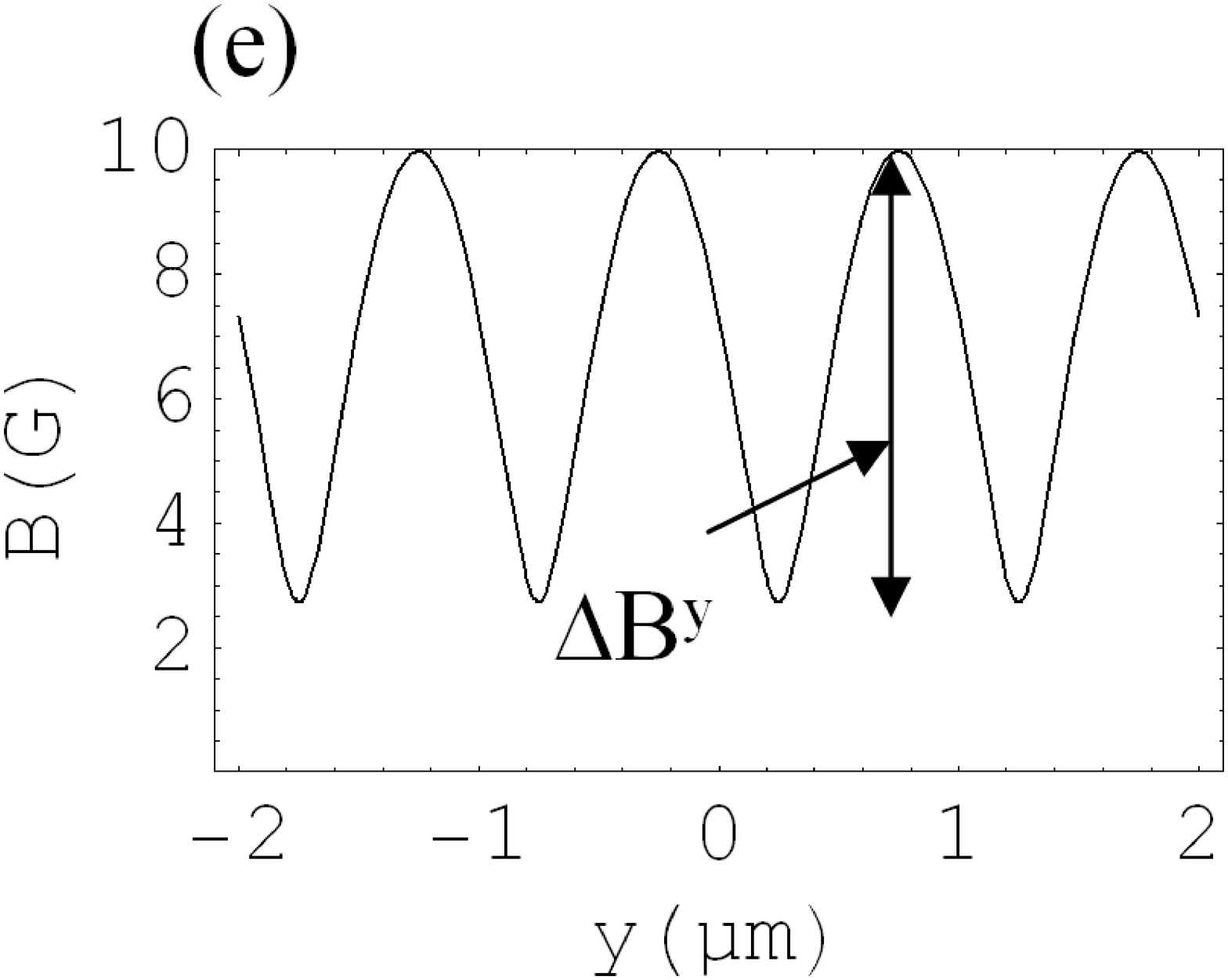} &
\hspace{.4cm}\includegraphics[angle=90,width=3.62cm,height=3.2cm]{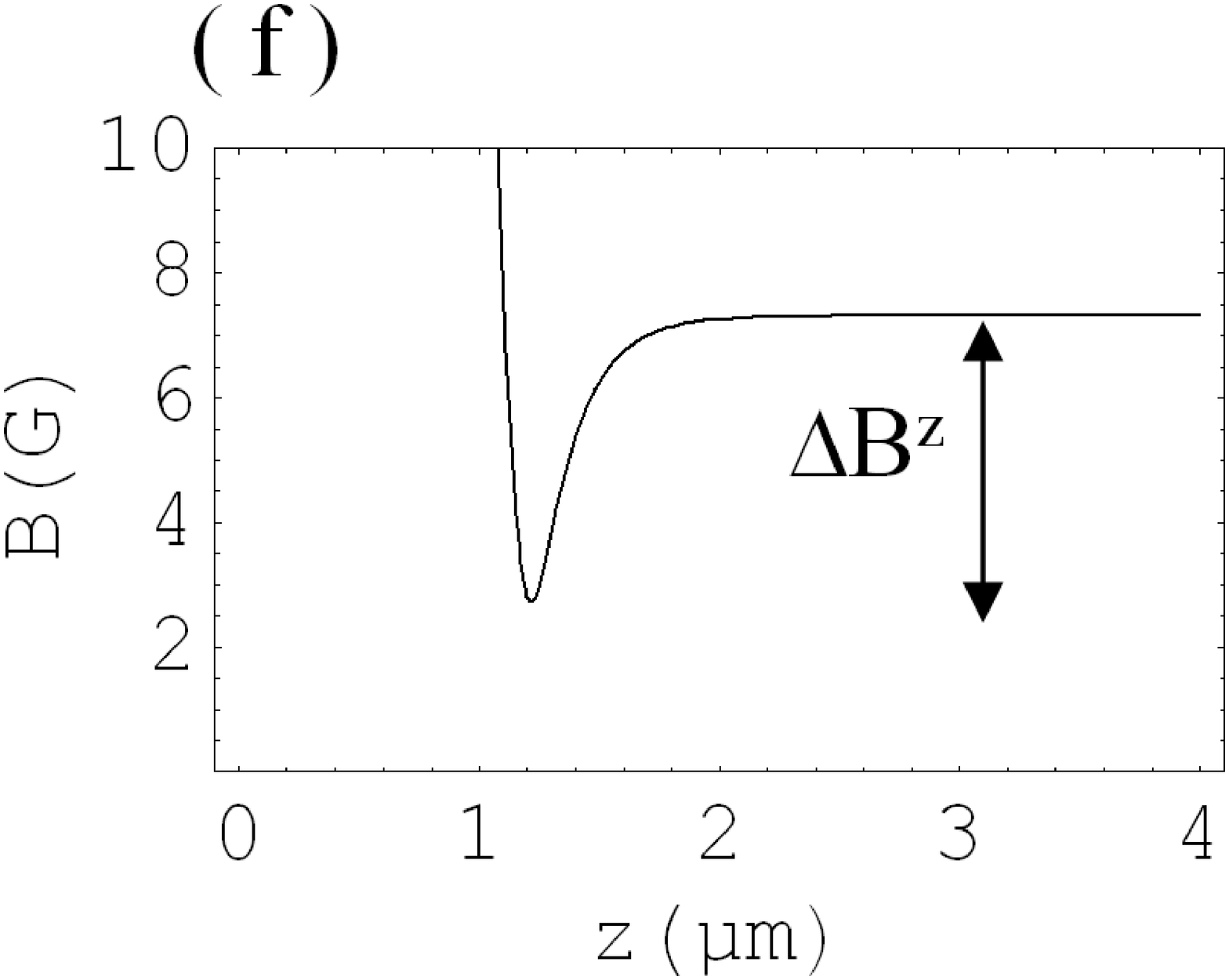}\\
\hspace{-0.1cm}\includegraphics[angle=90,width=3.97cm,height=3.2cm]{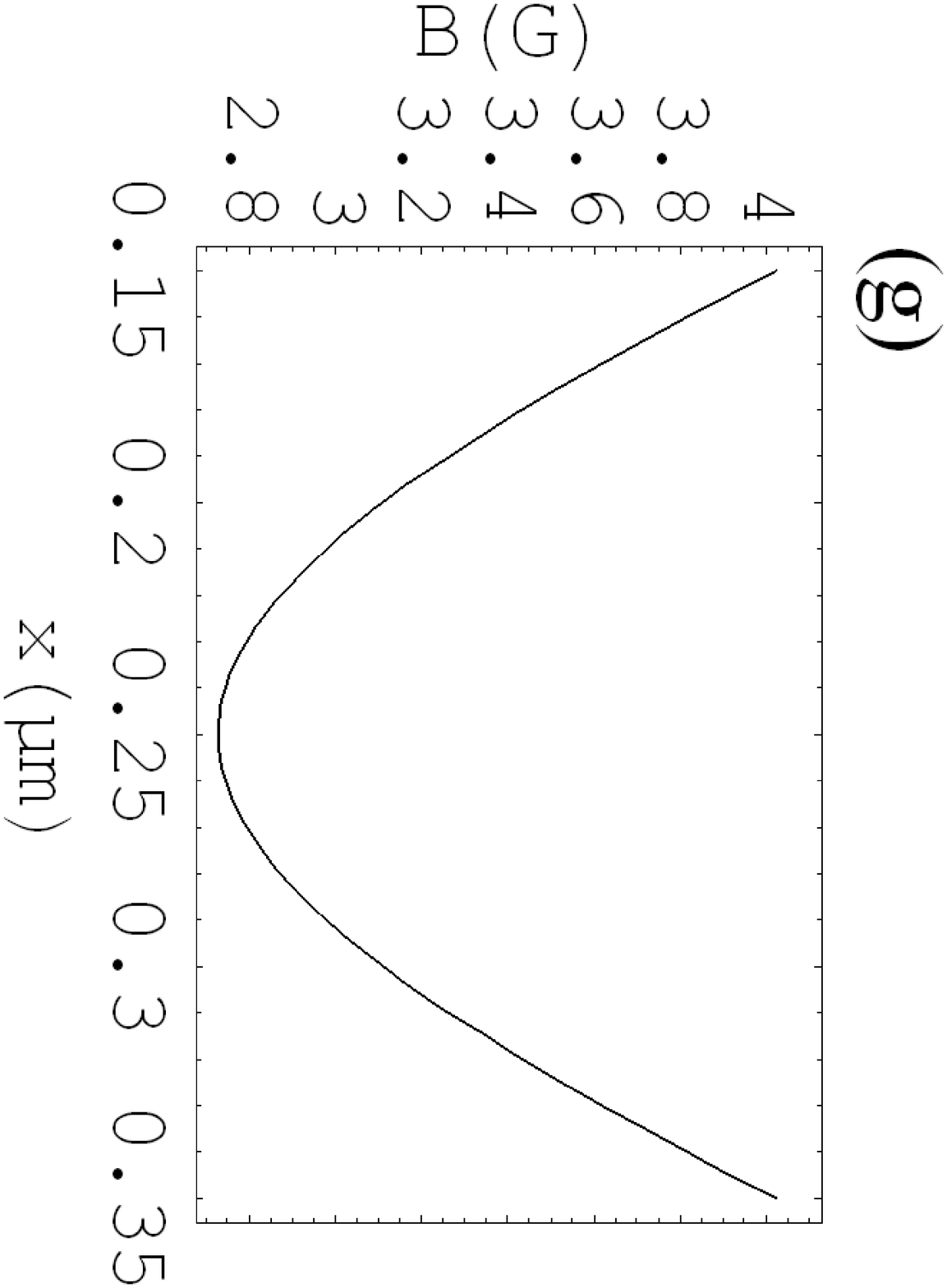}&
\hspace{-0.4cm}\includegraphics[angle=90,width=3.97cm,height=3.2cm]{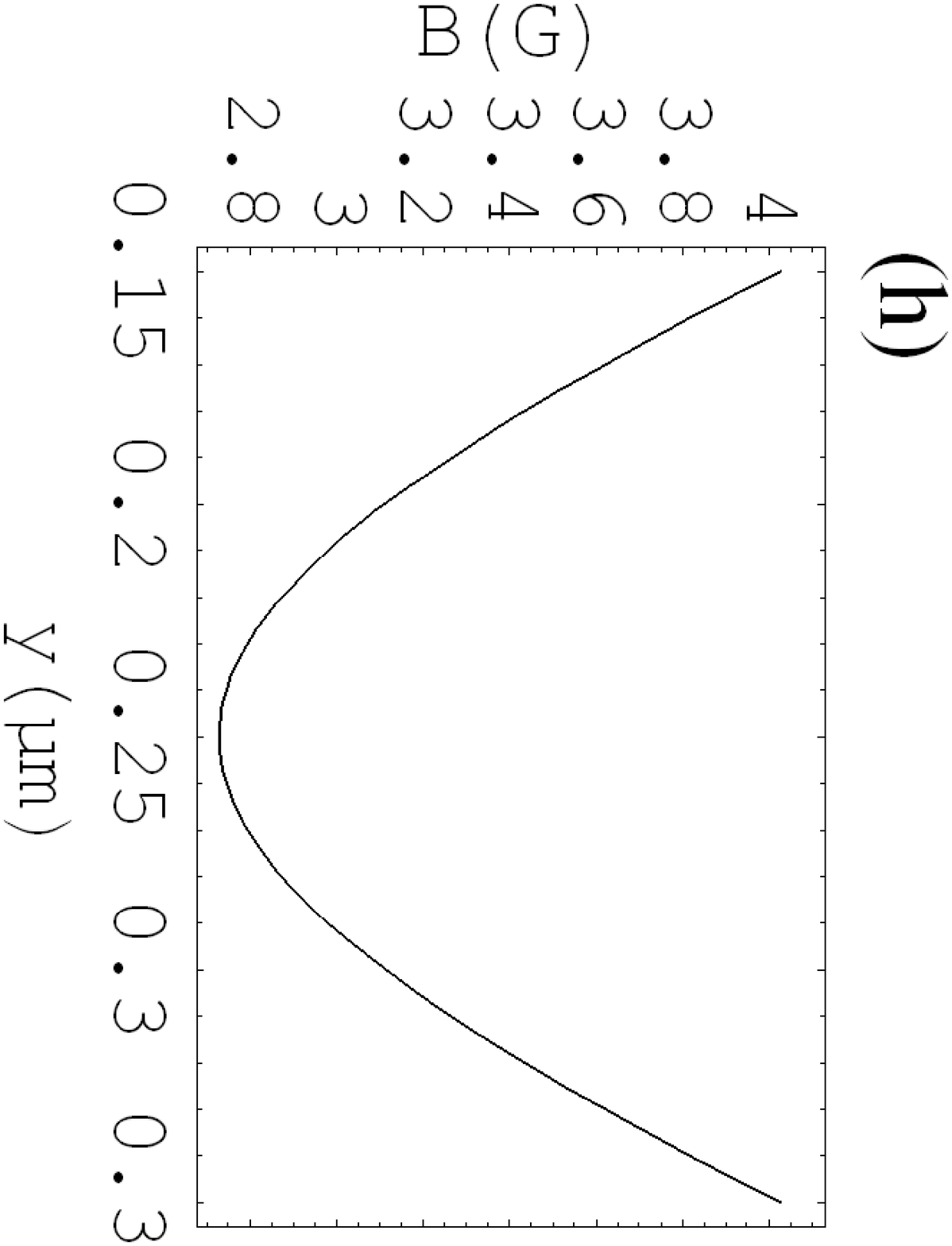}&
\hspace{-.9cm}\includegraphics[angle=90,width=3.97cm,height=3.2cm]{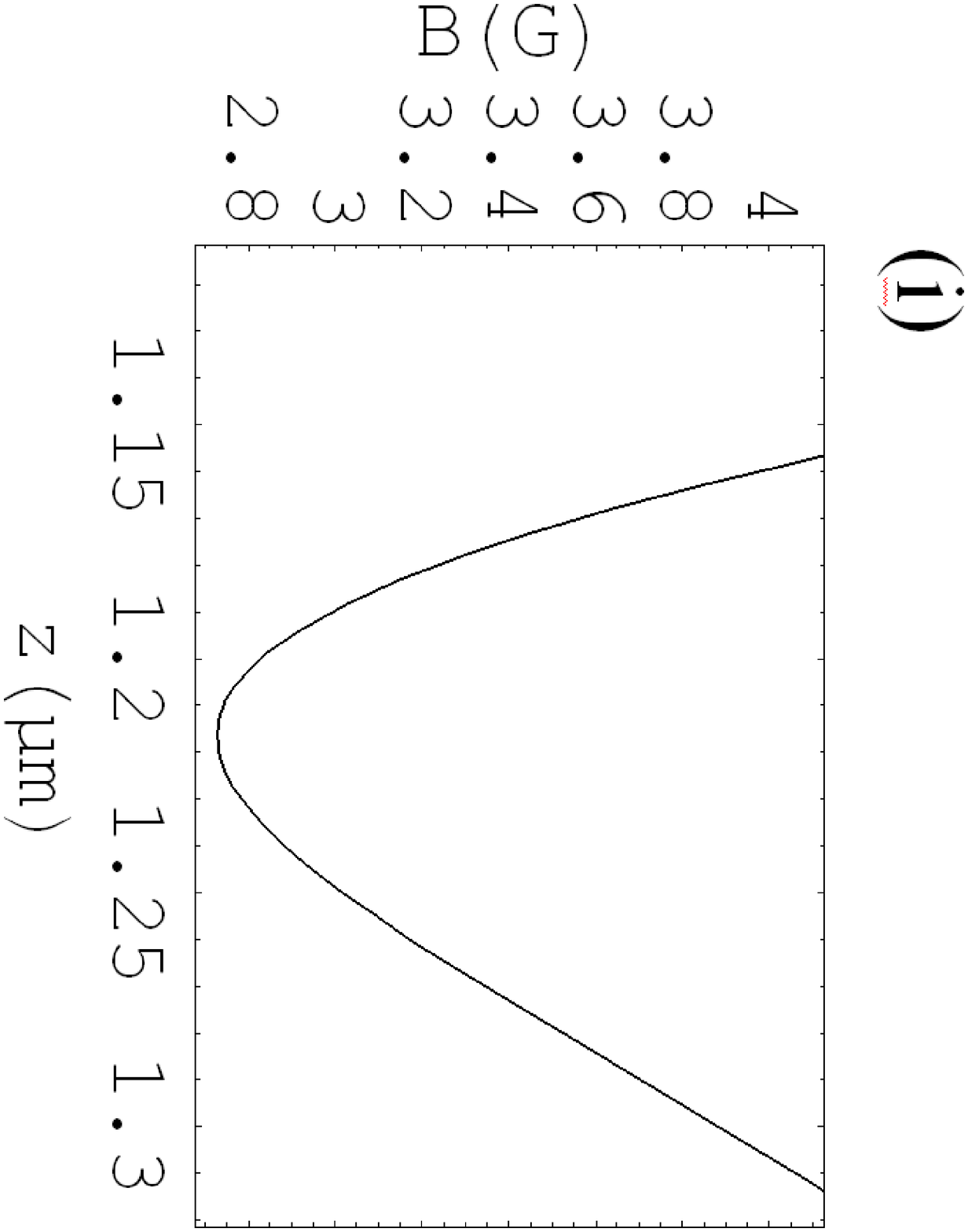}
\end{array}$
\caption{Magnetic field produced by a periodic array consisting of
two crossed layers of parallel rectangular magnets. (a) Near the
edge in the $x$ direction, (b) near the edge in the $y$-direction,
and (c) in the $z$ direction, for parameters given in \tref{table1}, column
3. Curves (d)-(f) show the  magnetic field in the central region of
the lattice with the abscissa expanded.  Curves (g)-(i) show the
magnetic field in the central region of the lattice plotted (g)
along a line ($y = y_{min}$, $z = z_{min}$) parallel to the
$x$-axis, (h) along a line ($x=x_{min}$, $z=z_{min}$) parallel to
the $y$-axis, and (i) along a line ($x=x_{min}$, $y=y_{min}$)
parallel to the $z$-axis. } \label{figure3}
\end{center}
\end{figure}
Non-zero potential
minima with $B_{min} =|B_{1x}| = 20 \hspace{.1cm}G$  appear in the potentials at
$z_{min} = 0.540 \hspace{.1cm}\mu m$. This configuration may be used as a 1D
periodic lattice of 2D magnetic traps or waveguides for slowly
moving atoms. Relatively large values of $B_{1x}$ and $B_{1y}$ were
used in this calculation so that the magnetic traps are sufficiently
deep to be seen in the contour plot. In practice smaller values
would normally be used. For example, if we use $B_{1x} =- 2.73 \hspace{.1cm}G$,
$B_{1y} =-0.22 \hspace{.1cm}G$ and $B_{1z} =- 0.09 \hspace{.1cm}G$, non-zero potential
minima with $B_{min} = |B_{1x}|=2.73 \hspace{.1cm}G$ appear in the potentials at $z_{min}
= 1.216 \hspace{.1cm}\mu m$.

\subsection{Two crossed layers of finite periodic arrays of magnets with
bias fields}{\label{sec3.2}}
\Fref{figure3} shows a numerical calculation of the magnetic field
versus distance in the $x$-, $y$- and $z$-directions of a {\it finite} array
consisting of two crossed layers of periodic arrays of parallel
rectangular magnets. The parameters are $n_r = 1001$, $a = 1 \hspace{.1cm}\mu m$,
$s = 0.1 \hspace{.1cm}\mu m$, $t_1 = 0.322\hspace{.1cm}\mu  m$ (bottom layer), $t_2 = 0.083\hspace{.1cm}
\mu m$ (top layer), $4\pi M_z = 3.8 \hspace{.1cm}kG$, $B_{1x} = - 4.08 \hspace{.1cm}G$,
$B_{1y} = - 6.05 \hspace{.1cm}G$ and $B_{1z} = - 0.69 \hspace{.1cm}G$ (see \tref{table1}).
\begin{figure}[tbp]
\begin{center}
$\begin{array}{cc}
\includegraphics[angle=90,width=6cm]{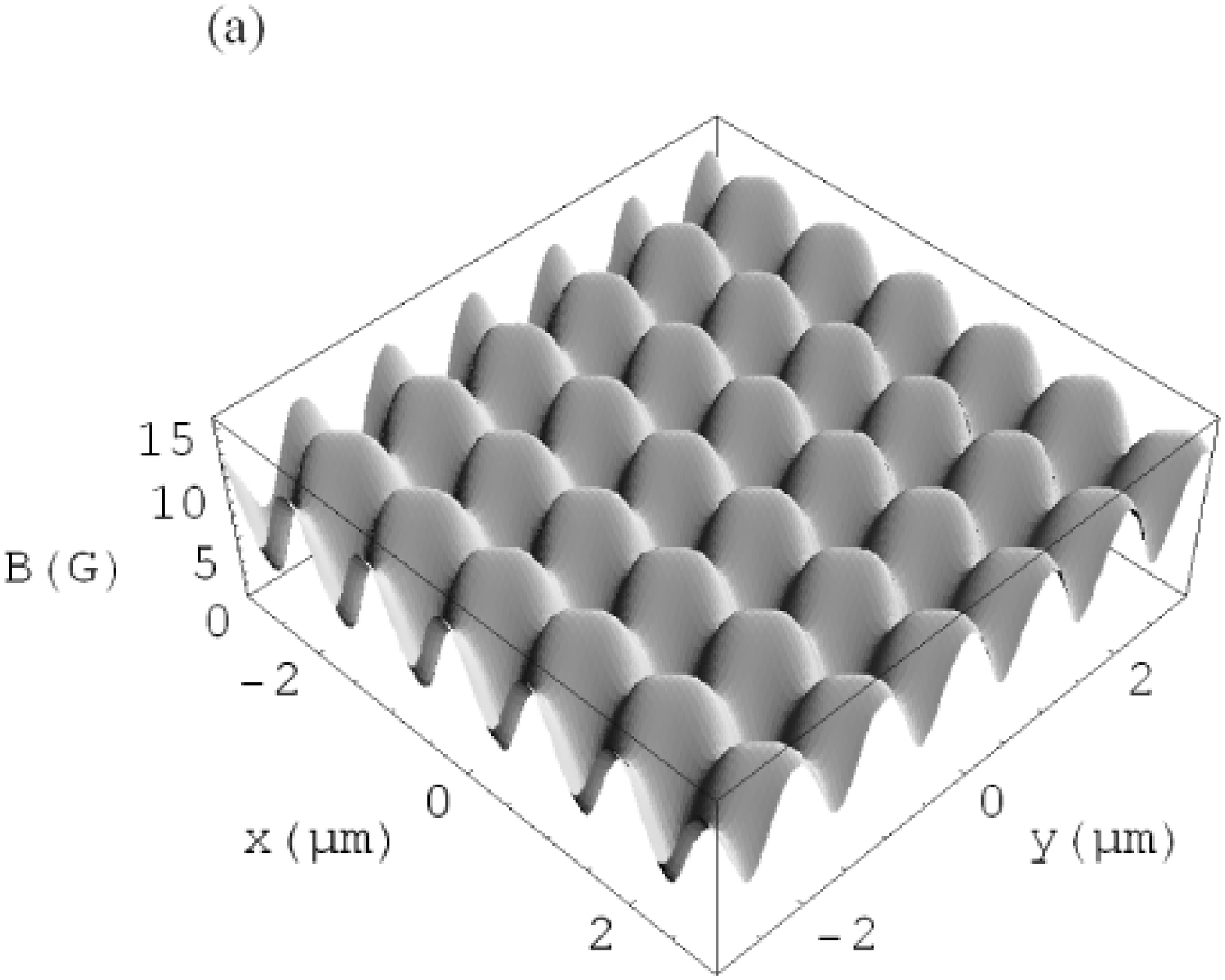} &
\includegraphics[width=4.5cm]{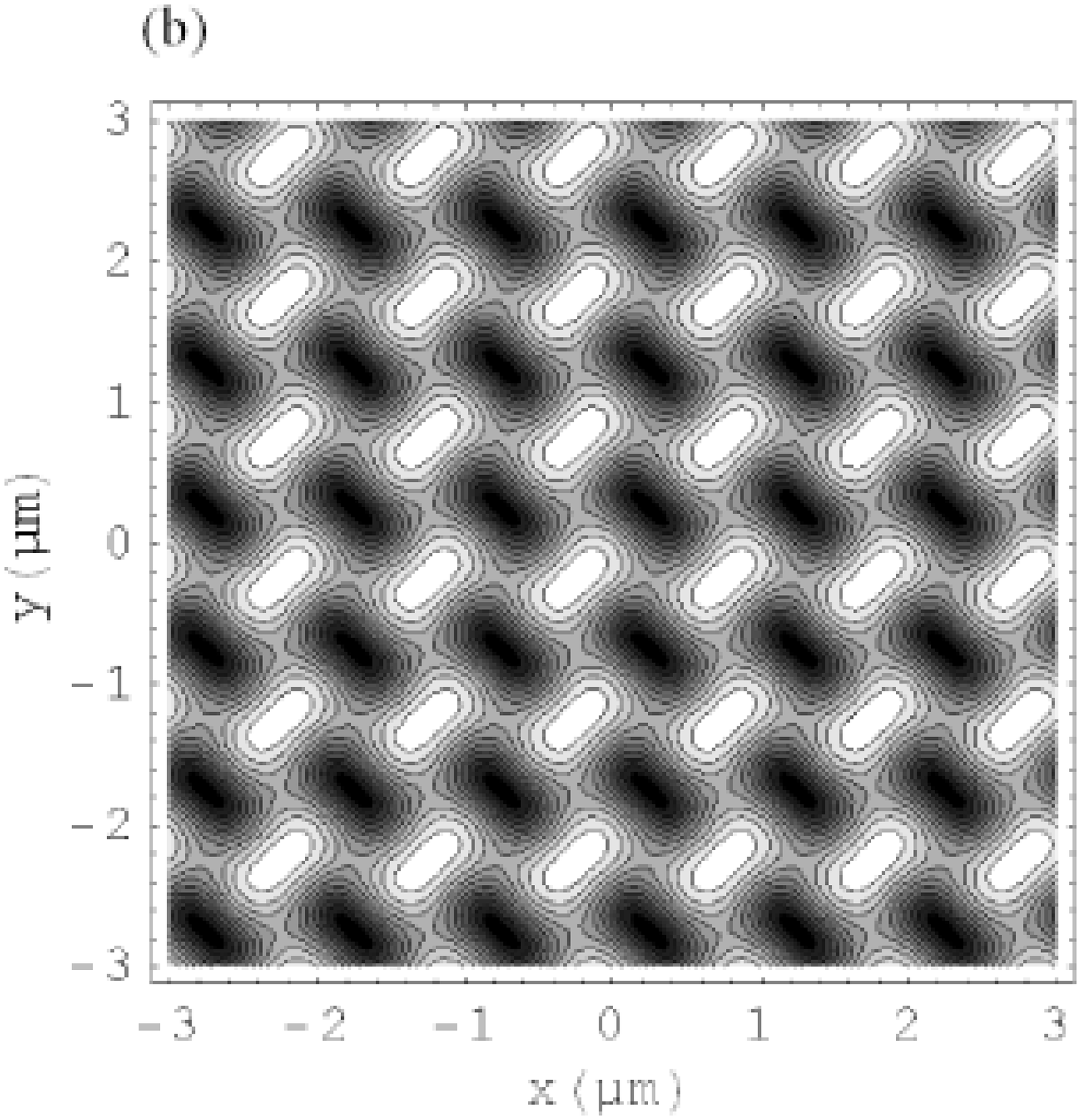}\end{array}$
\caption{Magnetic field in the plane $z=z_{min}$ for a periodic
array consisting of two crossed layers of parallel rectangular
magnets. (a) $3D$ plot, and (b) contour plot, for parameters given in \tref{table1}, column 3.}\label{figure4}
\end{center}
\end{figure}
\begin{table}[hb]
\caption{\label{table2}Parameters calculated numerically using Radia (columns 3 and 6)
 and using the analytical formulae (columns 4 and 5) for the input parameters in
\tref{table1}
 for (1) two crossed layers of periodic arrays of rectangular magnets (\fref{figure2}), and
 (2) a single layer of square magnets with three thicknesses (\fref{figure7}).}
\hspace{-.15cm}
{\footnotesize
\begin{tabular}{@{}llllll}
 \br
&&\centre{3}{Config.(1)}&\centre{1}{Config.(2)}\\
\ns
&&\crule{3}&\\
&&&\centre{2}{Analytical}&\\
&&&\crule{2}&\\
Parameter&Definition&Numerical& $B_{1z} = 0$& $B_{1z}= -0.69 \hspace{.1cm}(G) $& Numerical\\
\mr
\verb   $x_{min}\hspace{.1cm}(\mu m)$       &  $x$ co-ordinate      &    $0.250  $   &   $0.250  $    &  $0.184$    &   $0.250 $   \\
\verb   $$       &of  potential     &    $$   &   $$    &  $$    &   $$   \\
\verb   $$       & minimum      &    $$   &   $$    &  $$    &   $$   \\
\verb   $y_{min}\hspace{.1cm}(\mu m)$       &  $y$ co-ordinate     &    $0.250  $   &   $0.250 $    &  $0.294 $    &   $0.250  $   \\
\verb   $$       & of  potential     &    $$   &   $$    &  $$    &   $$   \\
\verb   $$       & minimum      &    $$   &   $$    &  $$    &   $$   \\
\verb   $z_{min}\hspace{.1cm}(\mu m)$        &  $z$ co-ordinate      &    $1.216  $   &   $1.216 $    &  $1.206  $    &   $0.952  $   \\
\verb   $$       &of  potential     &    $$   &   $$    &  $$    &   $$   \\
\verb   $$       & minimum      &    $$   &   $$    &  $$    &   $$   \\
\verb   $d\hspace{.1cm}(\mu m)$      &  Distance of      &    $0.712 $   &   $0.712 $    &  $0.702 $    &   $0.732  $   \\
\verb            & central minimum    &    $$   &   $$    &  $$    &   $$   \\
\verb            &  from  surface  &    $$   &   $$    &  $$    &   $$   \\
\verb   $B_{min}\hspace{.1cm}(G)$       &  Magnetic field at     &    $2.73 $   &   $2.73$    &  $2.64 $    &   $1.10 $   \\
\verb            &   potential minimum   &    $$   &   $$    &  $$    &   $$   \\
\verb   $\frac{\partial^2 B}{\partial x^2}\hspace{.1cm}({G\over{cm}^2})$       &   Curvature of $B$    &    $3.32 \times 10^{10}$   &   $3.31 \times 10^{10}$    &  $3.77 \times 10^{10} $    &   $7.52 \times 10^{10} $   \\
\verb   $$       & along $x$     &    $$   &   $$    &  $$    &   $$   \\
\verb   $\frac{\partial^2 B}{\partial y^2}\hspace{.1cm}({G\over{cm}^2})$       &   Curvature of $B$   &    $3.32 \times 10^{10} $   &   $3.32 \times 10^{10} $    &  $3.77 \times 10^{10} $    &   $7.52 \times 10^{10} $   \\
\verb   $$       &  along $y$     &    $$   &   $$    &  $$    &   $$   \\
\verb   $\frac{\partial^2 B}{\partial z^2}\hspace{.1cm}({G\over{cm}^2})$       &   Curvature of $B$    &    $6.64 \times 10^{10} $   &   $6.64 \times 10^{10} $    &  $6.98 \times 10^{10} $    &   $1.50 \times 10^{11} $   \\
\verb   $$       &   along $z$   &    $$   &   $$    &  $$    &   $$   \\
\verb   $\Delta B^x\hspace{.1cm}(G)$       &  Magnetic barrier     &    $7.22$   &   $7.22$    &  $8.01$    &   $8.10$   \\
\verb   $$       &   height along $x$   &    $$   &   $$    &  $$    &   $$   \\
\verb   $\Delta B^y\hspace{.1cm}(G)$       &  Magnetic barrier     &    $7.23$   &   $7.23$    &  $7.78$    &   $8.09$   \\
\verb   $$       &   height along $y$   &    $$   &   $$    &  $$    &   $$   \\
\verb   $\Delta B^z\hspace{.1cm}(G)$       &  Magnetic barrier     &    $4.57$   &   $4.57$    &  $4.69$    &   $5.45$   \\
\verb   $$       &   height along $z$   &    $$   &   $$    &  $$    &   $$   \\
\br
\end{tabular}
}
\end{table}
\begin{table}[hb]
 \caption{\label{table3}Parameters calculated numerically using Radia for input parameters in
 \tref{table1} for ${}^{87}{\rm Rb}$  $F=2$, $m_F=+2$ atoms trapped in a magnetic lattice consisting of
 (1) two crossed layers of periodic arrays of rectangular magnets (\fref{figure2}), and
 (2) a single layer of square magnets with three thicknesses (\fref{figure7}).}
\hspace{.55cm}
{\footnotesize
{\footnotesize
\begin{tabular}{@{}llll}
\br
Parameter&Definition&Configuration (1)&Configuration (2)\\
\mr
\verb   $U_{min}/k_B \hspace{0.1cm}(\mu { K})$    & Energy of minimum of potential          &  $183 $     &   $74  $                 \\
\verb   $\Delta U^x/k_B \hspace{0.1cm}(\mu { K})$    & Potential barrier height along $x$   &  $485 $     &   $544 $                 \\
\verb   $\Delta U^y/k_B \hspace{0.1cm}(\mu { K})$    & Potential barrier height along $y$   &  $486 $     &   $544 $                 \\
\verb   $\Delta U^z/k_B \hspace{0.1cm}(\mu { K})$    & Potential barrier height along $z$   &  $307 $     &   $383 $                 \\
\verb   $\omega_x/ 2\pi \hspace{0.1cm}({ kH}z)$   &  Trap frequency along $x$            &  $232  $       &   $350 $                \\
\verb   $\omega_y/ 2\pi \hspace{0.1cm}({ kH}z)$   &  Trap frequency along $y$            &  $233  $       &   $350 $                \\
\verb   $\omega_z/ 2\pi \hspace{0.1cm}({ kH}z)$   &  Trap frequency along $z$            &  $329  $       &   $494 $                 \\
\verb   $\hbar\omega_x/k_B \hspace{0.1cm}(\mu { K})$ &  Level spacing along $x$             &  $11$      &   $17 $                  \\
\verb   $\hbar\omega_y/k_B \hspace{0.1cm}(\mu { K})$ &  Level spacing along $y$             &  $11 $      &   $17 $                  \\
\verb   $\hbar\omega_z/k_B \hspace{0.1cm}(\mu { K})$ &  Level spacing along $z$             &  $16 $      &   $24 $                  \\
\br
\end{tabular}
}
}
\end{table}
The values of $B_{1x}$, $B_{1y}$, $s$, $t_1$ and $t_2$ satisfy the
condition for a symmetrical {\it infinite} magnetic lattice with two
bias fields [equations \eref{E15a} or \eref{E15b}], while the value
of $B_{1z}$ was adjusted to compensate for asymmetry introduced into
the lattice by `end-effects' associated with the finite number
(1001) of magnets in the array (see \sref{sec2.2.3}).  The peaks at
the ends of the array  [\fref{figure3}(a) and (b)] also arise from
end-effects. The potential of the microtraps is close to harmonic in
the region near the bottom of the traps [\fref{figure3} (g)-(i)].

\Fref{figure4} shows a 3D plot and a contour plot  of the magnetic
field in the plane $z = z_{min}$ for the two crossed-layer structure
of parallel magnets. The parameters used in this calculation
(\tref{table1}, column 3) lead to $(n_r - 1)^2 = 10^6$  magnetic
microtraps with $B_{min} = 2.73 \hspace{.1cm}G$ at the {\it central }
minimum, which is located at $d = 0.712 \hspace{.1cm} \mu m $ from the top
surface, and magnetic field barrier heights $\Delta B^x = B^x_{max}
- B_{min} = \Delta B^y = 7.23\hspace{.1cm} G$ and  $\Delta B^z= 4.57 \hspace{.1cm}G$. The
potential minima in the outer regions of the array have slightly
different values for the barrier heights due to end-effects.  The
various quantities determined from this calculation are listed in
\tref{table2}, column 3, along with values determined using the
analytical expressions for an infinite symmetrical magnetic lattice
with $B_{1z} = 0$ (column 4) and $B_{1z} = - 0.69\hspace{.1cm} G$ (column 5).
The values determined from the numerical calculations for the finite
lattice with $B_{1z} =- 0.69 \hspace{.1cm}G$ are in excellent agreement with
those determined from the analytical expressions for the infinite,
symmetrical lattice with $B_{1z} = 0$.

The height of the potential barrier in each direction $i \hspace{.1cm}(i = x, y,
z)$ for the trapped ultracold atoms is related to the magnetic field
barrier height by
\begin{equation}\label{E28}
\Delta U^i= U^i_{max} - U_{min} = m_{{}_F}g_{{}_F}\mu_{{}_B}\Delta B^i
\end{equation}
The calculated parameters for ${}^{87}{\rm Rb}$ atoms in the low
magnetic field-seeking $F = 2$, $m_{{}_F} = +2$ state are listed in
column 3 of \tref{table3}. The potential barrier heights are  $\Delta U^x =485 \hspace{.1cm}\mu
K$, $\Delta U^y = 486 \hspace{.1cm}\mu
K$ and $\Delta U^z = 307\hspace{.1cm}\mu K$, and the trap frequencies are $\omega_x= 2\pi\times 232 \hspace{.1cm}kHz$,
$\omega_y= 2\pi\times 233 \hspace{.1cm}kHz$ and $\omega_z = 2\pi \times 329 \hspace{.1cm}kHz$.
The trap frequencies may be scaled down, if necessary, by reducing $c_0 = {B_{0x}/B_{0y}}$ or $B_{1x}$ and $B_{1y}$ [equation\eref{E22}].
\begin{figure}[tbp]
\vspace{.2cm}
\begin{center}
\includegraphics[angle=90,width=5.4cm,height=3.4cm]{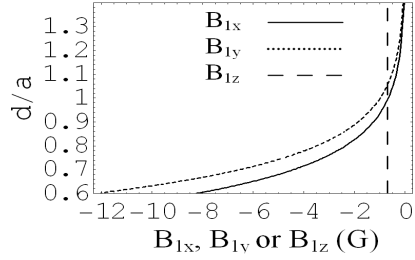}\label{figure5}
\caption{Distance of the central minimum from the surface $d =
z_{min} - (t_1 + t_2 + s)$ (in units of $a$) versus bias fields
$B_{1x}$, $B_{1y}$ or $B_{1z}$, for the parameters given in \tref{table1},
column 3.}
\end{center}
\end{figure}
\begin{figure}[tbp]
\begin{center}
$\begin{array}{ccc}
\includegraphics[angle=90,width=4cm,height=3.4cm]{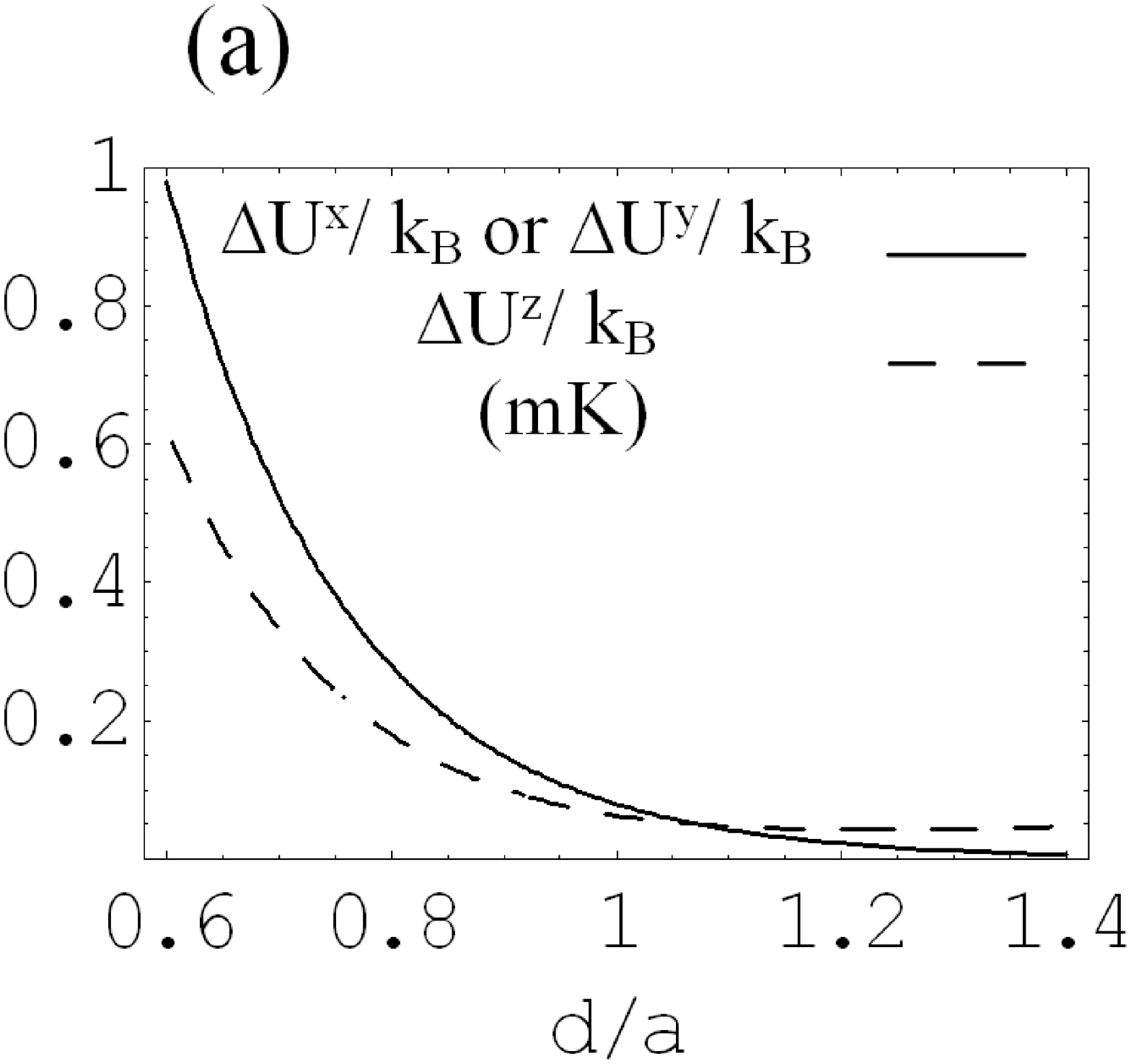} &
\includegraphics[angle=90,width=4cm,height=3.42cm]{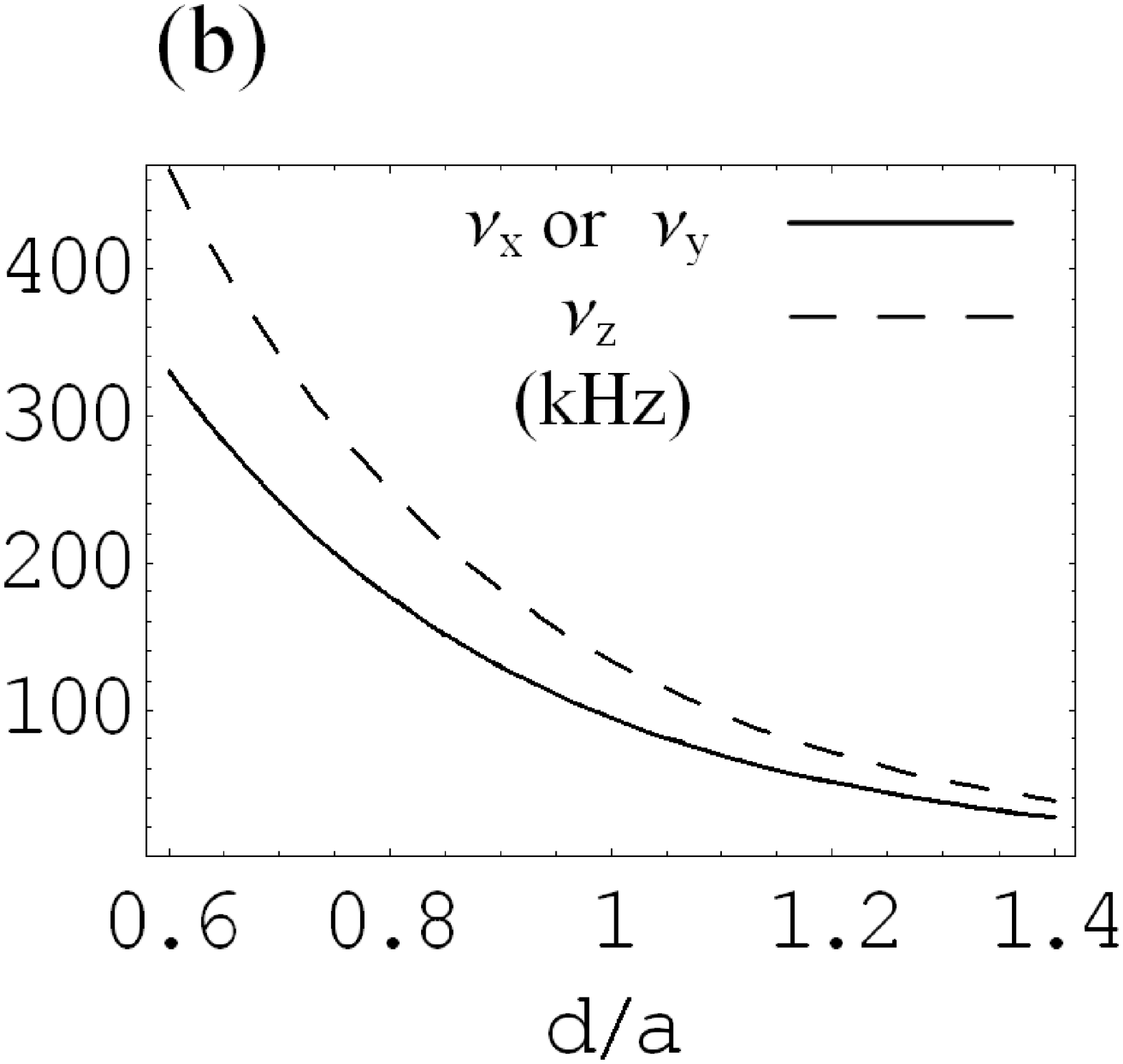} &
\includegraphics[angle=90,width=4cm,height=3.5cm]{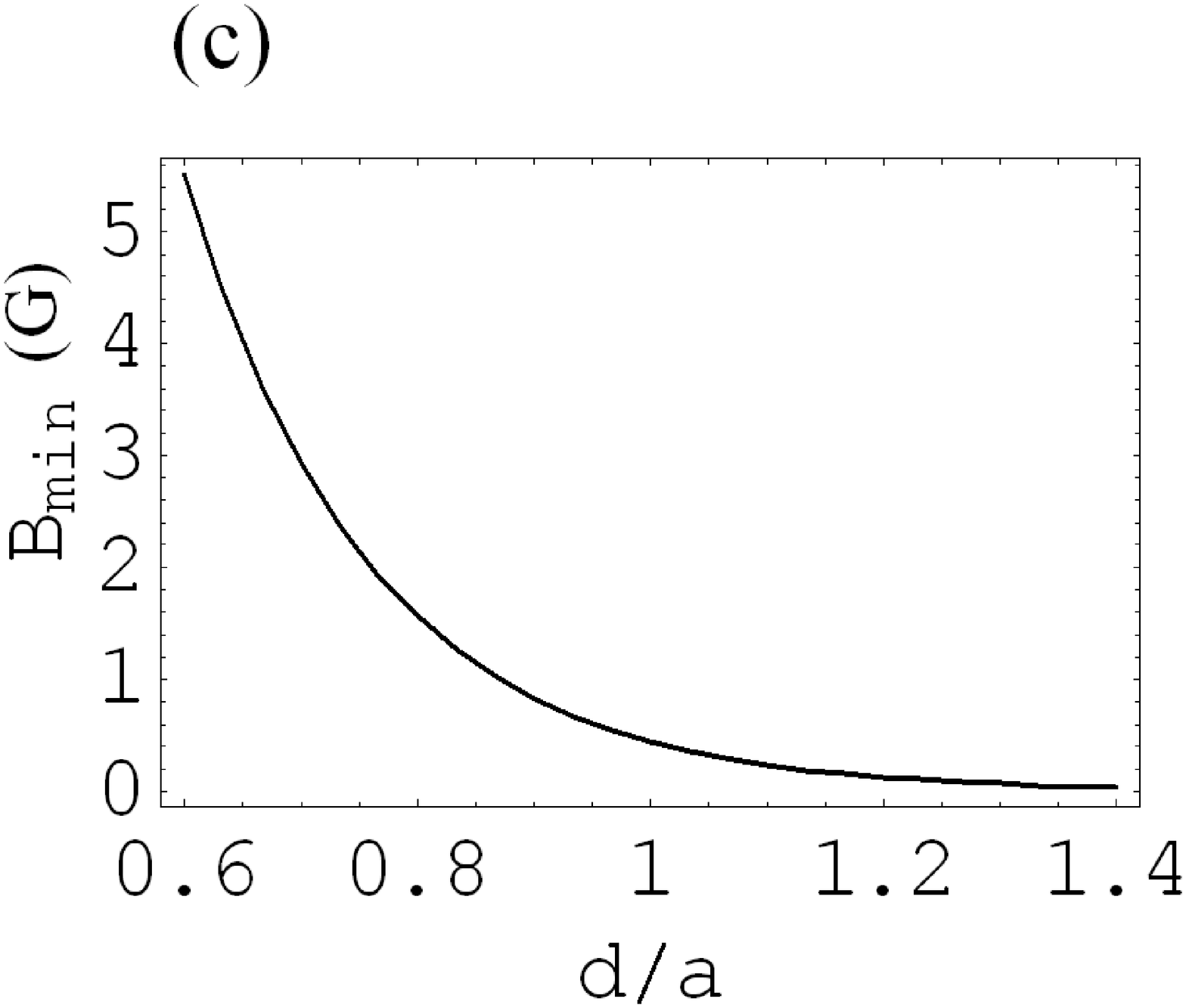}
\end{array}$
\caption{Potential barriers (a) and trap frequencies (b) in the
$x$-, $y$- and $z$-directions, and $B_{min}$ (c) for the central
minimum, as a function of $d/a$, for the parameters given in \tref{table1}, column 3. }
\label{figure6}
\end{center}
\end{figure}

Figure 5 shows how the distance of the central minimum from the
surface, $d = z_{min} - (t_1 + t_2 + s)$, expressed in units of the
period $a$, decreases with increasing strength of the bias
magnetic fields $B_{1x}$ and  $B_{1y}$.  \Fref{figure6}(a), (b) and (c) show the
exponential increase in magnetic field minimum,
potential barrier heights and trap frequencies  with decreasing distance $d$ of the minimum from the surface.

Figures 5 and \ref{figure6} illustrate how the magnetic field minimum, potential barrier heights and the trap frequencies of the microtraps
in the magnetic lattice may be controlled by varying the bias magnetic fields $B_{1x}$ and  $B_{1y}$ to move the microtraps closer to or
further from the magnetic array. We note that $B_{1x}$ and  $B_{1y}$ should be varied simultaneously, according to \eref{E15a}, while $B_{1z}$ is constant.

\subsection{Single layer of periodic arrays of permanent magnets}\label{sec3.3}
A second configuration (\fref{figure7}), which leads to
qualitatively similar 2D magnetic lattices to the two crossed layers
of parallel rectangular magnets, but which may be easier to fabricate, consists
of a single layer of square-shaped magnets having three different
thicknesses plus bias fields in the $x$-, $y$- and $z$-directions,
where the thickness of the thickest magnet $t_3 = t_1 + t_2$.
 The parameters used
in the numerical calculation are listed in column 4 of \tref{table1}
and the quantities determined from this calculation are summarized
in the final columns of tables \ref{table2} and \ref{table3}.
\begin{figure}[tbp]
\vspace{.3cm} \nonumber
\begin{center}
$\begin{array}{ccc}
\includegraphics[angle=0,width=3.8cm,height=3.8cm]{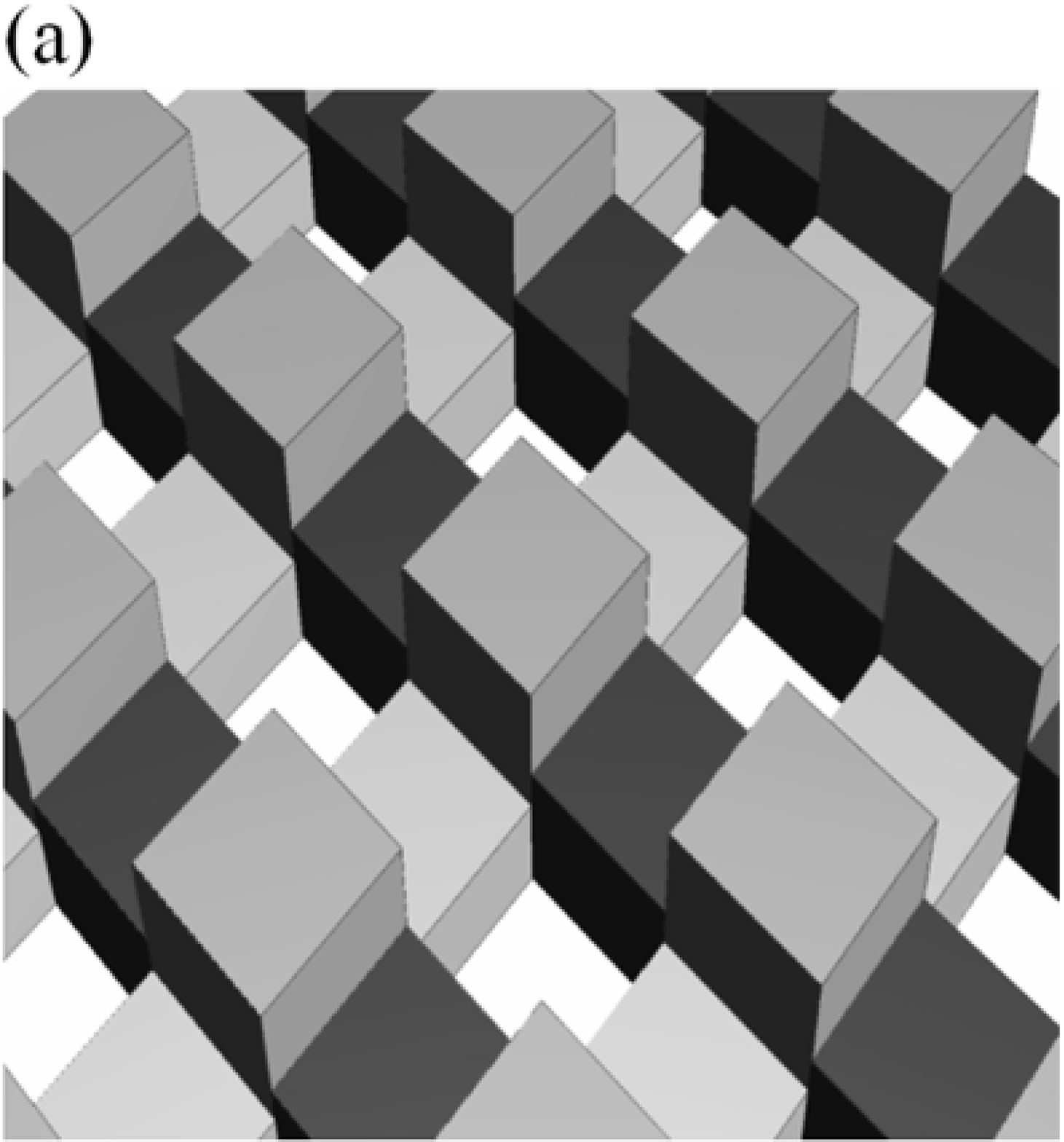} &
\includegraphics[angle=0,width=3.8cm,height=3.93cm]{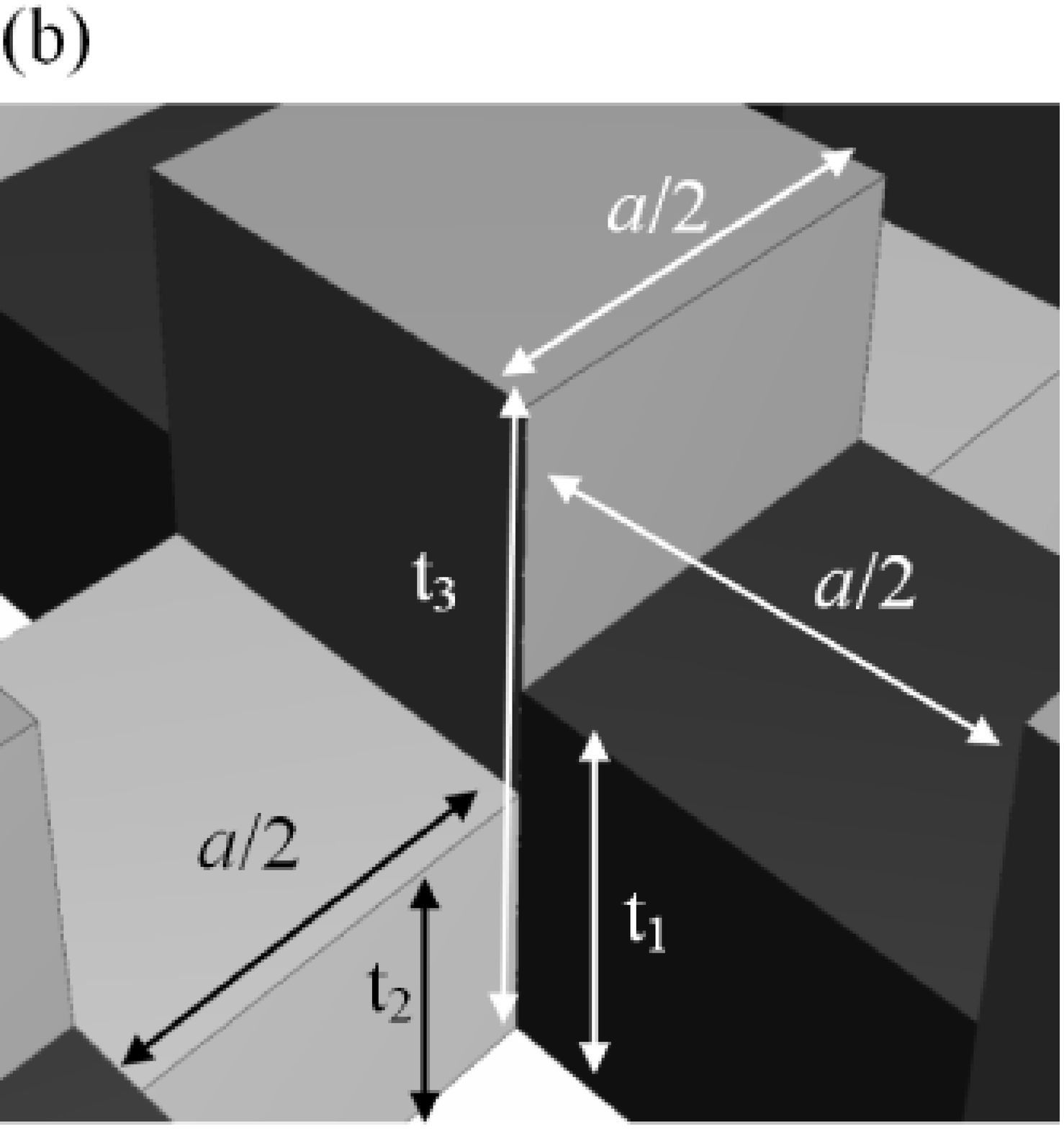}&
{\hspace{-.3cm}{\includegraphics[angle=0,width=4.cm,height=4.01cm]{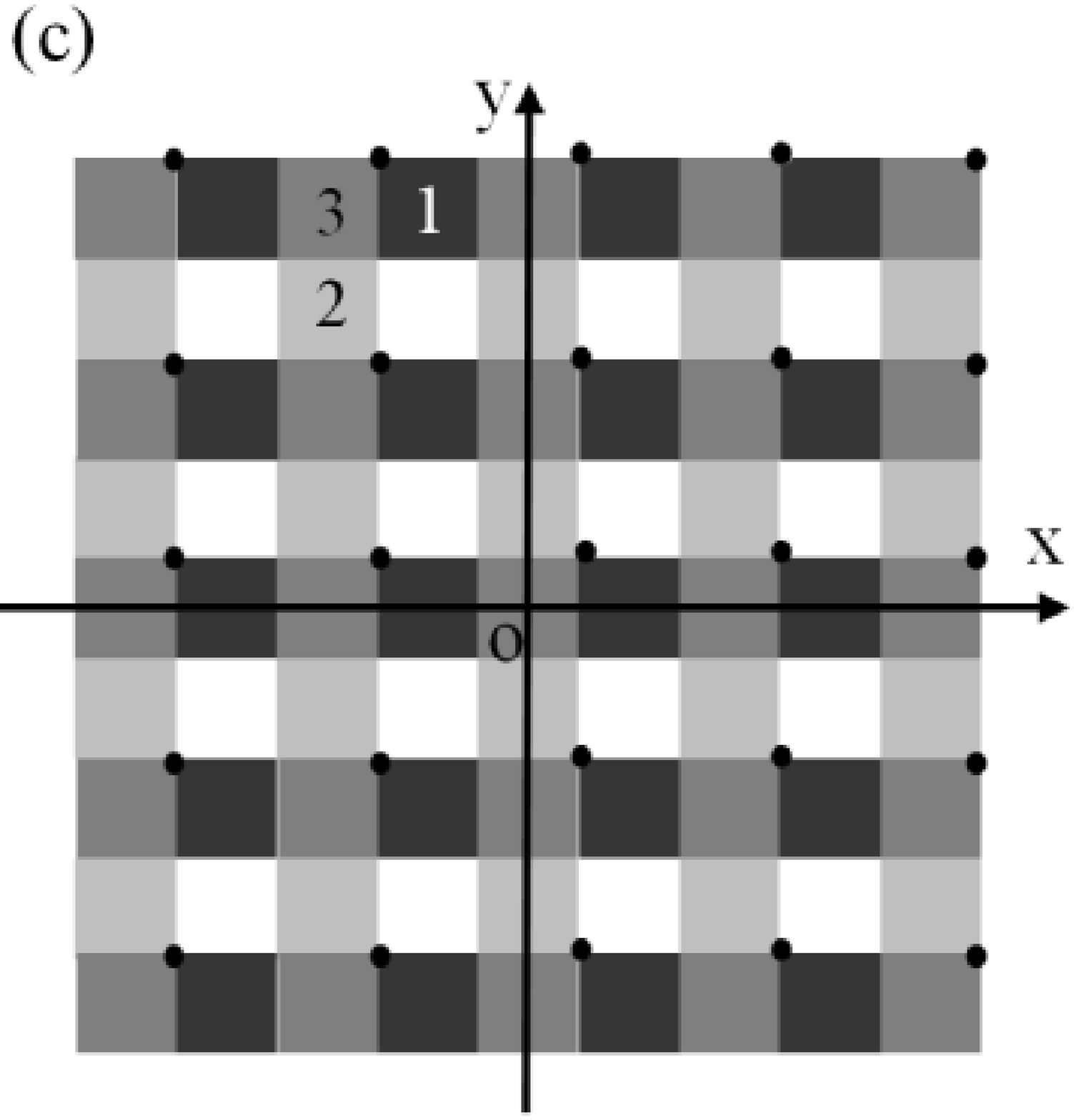}}}
\end{array}$
\end{center}
\end{figure}
\begin{figure}[tbp]
\vspace{-.3cm}
\begin{center}
$\begin{array}{cc}
\includegraphics[angle=90,width=3.8cm,height=3.cm]{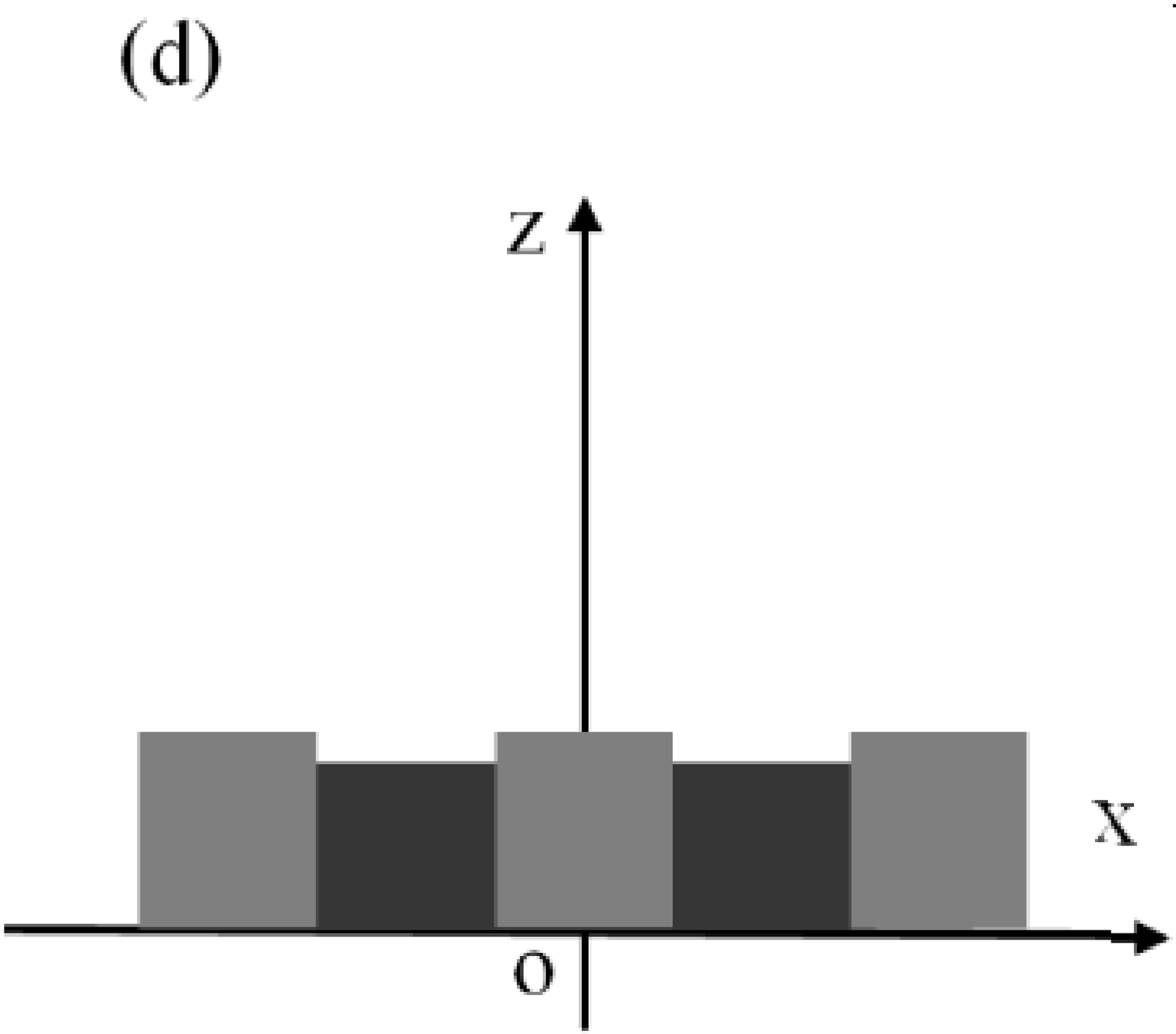}&
\includegraphics[angle=90,width=3.8cm,height=3.cm]{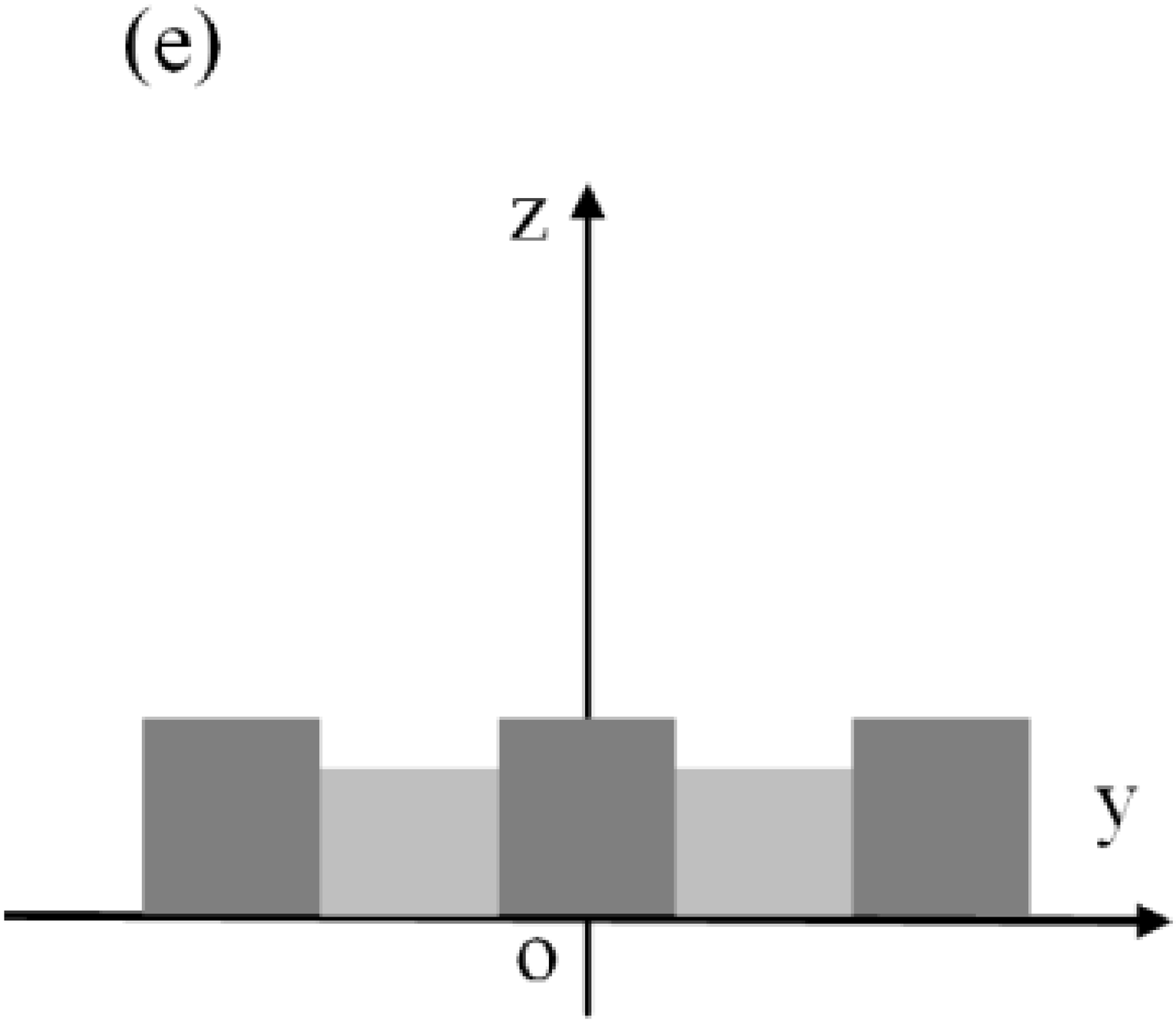}
\end{array}$
\caption{Periodic array consisting of one layer of square magnets
with three different thicknesses.  In (c) the locations of the
central minima in the $x$-$y$ plane are shown for a symmetrical
magnetic lattice with parameters given in \tref{table1}, column 4.
}\label{figure7}
\end{center}
\end{figure}
\section{Discussion and summary}\label{sec4}
Using analytical expressions and numerical calculations we have shown that periodic arrays of permanent
magnetic films plus bias magnetic fields can lead to 2D or 1D magnetic lattices of microtraps having
{\it non-zero} potential minima and controllable trap depth. Two configurations have been found that
lead to 2D magnetic lattices with non-zero potential minima: the first consists of two crossed layers
 of periodic arrays of parallel rectangular magnets plus bias fields and the second consists of a single
layer of square-shaped magnets having three different thicknesses
plus bias fields. These configurations lead to a {\it symmetrical}
magnetic lattice with equal barrier heights in the $x$- and
$y$-directions in the plane of the array if the bias fields $B_{1x}$
and $B_{1y}$ maintain a fixed relationship [given by \eref{E15a} or
\eref{E15b} for the crossed array configuration]
 and if the bias field $B_{1z}$ normal to the array is adjusted to compensate for asymmetry introduced by
 the finite size of the array.  In some experiments, e.g., in studies of quantum tunnelling between
lattice sites, it may be useful to be able to vary $B_{1x}$ and $B_{1y}$ independently in order to
vary the relative barrier heights in the $x$- and  $y$-directions.

For arrays with micron-scale periodicity, the magnetic microtraps can have very large trap depths
($\sim \hspace{-.05cm}0.5 \hspace{.1cm}mK$ for the parameters chosen for the 2D lattice), allowing relatively warm atoms to be trapped,
and very tight confinement.  The barrier heights of the microtraps can be controlled by varying the
bias fields $B_{1x}$ and $B_{1y}$ in the plane of the array, to move the traps either closer to or
further from the surface.  The numerical calculations for the crossed array configuration were
performed for a $1 \hspace{.1cm}mm  \times 1 \hspace{.1cm}mm$ magnetic lattice with period $a = 1 \hspace{.1cm}\mu m$, giving $10^6$ lattice sites.
It should be straight-forward to scale up the magnetic lattice, for example, to a $1 \hspace{.1cm}cm  \times 1 \hspace{.1cm}cm$
lattice with period $a = 1 \hspace{.1cm}\mu m$, giving $10^8$ lattice sites.

The permanent magnetic lattice configurations considered here should be suitable
for trapping and manipulating small clouds of ultracold atoms prepared in
low magnetic field-seeking states, including Bose-Einstein condensates and
ultracold Fermi gases.  A cloud of ultracold atoms could be loaded into
the permanent magnetic lattice using, for example, a hybrid magnetic field structure
comprising a current-carrying `U' quadrupole trap and a `Z' Ioffe-Pritchard trap
located beneath the permanent magnetic array on the atom chip~\cite{Hall}.
 Such a hybrid structure should allow ultracold atoms to be initially loaded from
a mirror MOT into the `U' surface MOT, then into the `Z' magnetic trap, and finally into
the 2D magnetic lattice.

By using a small period ($a\hspace{-.05cm}\sim 1\hspace{.1cm}\mu m$) and controlling the barrier height between
the microtraps it may be possible to perform quantum tunnelling experiments
such as the BEC superfluid to Mott insulator transition~\cite{Greiner} in a 2D magnetic lattice.
Some of the challenges will include the ability to fabricate permanent magnetic
arrays with sufficiently smooth magnetic potentials and {\it equivalent} magnetic microtraps
and to mimimise the effects of the interaction of the ultracold atoms with the surface~\cite{Henkel,Folman,Jones,Harber,Lin,Scheel,Casimir}
in order to preserve quantum coherence of the atoms in the magnetic lattice.
For the magnetic lattices considered here, the potential minima are located about $1 \hspace{.1cm}\mu m$
from the surface.  At such distances the Casimir Polder force~\cite{Casimir}  can be significant~\cite{Lin},
leading to an attractive component that lowers the barrier height, and losses due to
thermally induced spin flips caused by interaction with the ambient temperature surface~\cite{Jones,Harber,Lin,Scheel,Casimir}
can be important.  However, spin-flip losses can be minimised by using magnetic films
whose thickness ($t \le  0.4 \hspace{.1cm}\mu m$ for the magnetic lattices considered here) is much less than
the skin depth and by use of suitable dielectric substrates with low electrical conductivity~\cite{Henkel}.
It should be possible to further reduce interactions with the surface, if necessary,
by moving the potential minima further from the surface by decreasing the bias magnetic
fields and/or marginally increasing the period of the magnetic lattice.
Use of the BEC to Mott insulator transition for ultracold atoms trapped in 2D magnetic
lattices could allow the preparation of a single qubit atom on each magnetic lattice site,
which is important for scalable quantum information processing.

\ack{ We thank Romain Pari\`{e}s for contributions to the early
calculations, and Shannon Whitlock, Brenton Hall, Bryan Dalton and Giovanni Modugno
for stimulating discussions.
This work is supported by a Swinburne University Strategic
Initiative Grant.  Saeed Ghanbari thanks the Iranian Government for financial
support.}

\end{document}